\documentclass[10pt,aip,apl,reprint,citeautoscript,superscriptaddress,nofootinbib,nobalancelastpage,floatfix]{revtex4-1}

\usepackage{siunitx}    
\usepackage{graphicx}         
\usepackage{fancyhdr}                 
\usepackage{amsmath}                  
\usepackage{bm}                       
\usepackage{xcolor}
\usepackage[colorlinks=true,
    citecolor=black,
    linkcolor=black,
    urlcolor=blue,
    pagebackref=false,
    breaklinks=true]{hyperref}
\usepackage{xr}
\usepackage{natbib}  

\usepackage{times}

\newcommand{\st}[1]{_{\text{#1}}}
\newcommand{\latin}[1]{#1}

\begin{document}



    \title{Vector Electric Field Measurement via Position-Modulated Kelvin Probe Force Microscopy}

    \author{Ryan P. Dwyer}
    \affiliation{Department of Chemistry and Chemical Biology, Cornell University, Ithaca, New York 14853, USA}

    \author{Louisa M. Smieska}
    \affiliation{Department of Chemistry and Chemical Biology, Cornell University, Ithaca, New York 14853, USA}

    \author{Ali Moeed Tirmzi}
    \affiliation{Department of Chemistry and Chemical Biology, Cornell University, Ithaca, New York 14853, USA}

    \author{John A. Marohn}
    \affiliation{Department of Chemistry and Chemical Biology, Cornell University, Ithaca, New York 14853, USA}


\begin{abstract}
High-quality spatially-resolved measurements of electric fields are critical to understanding charge injection, charge transport, and charge trapping in semiconducting materials.
Here, we report a variation of frequency-modulated Kelvin probe force microscopy (FM-KPFM) that enables spatially-resolved measurements of the electric field.
We measure electric field components along multiple directions simultaneously by employing position modulation and lock-in detection in addition to numeric differentiation of the surface potential.
We demonstrate the technique by recording linescans of the in-plane electric field vector in the vicinity of a patch of trapped charge in a DPh-BTBT organic field-effect transistor.
This technique is simple to implement and should be especially useful for studying electric fields in spatially inhomogeneous samples like organic transistors and photovoltaic blends.
\end{abstract}

\date{\today}

\maketitle
\thispagestyle{fancy}



In this letter we describe a simple modification of frequency-modulated Kelvin probe force microscopy\cite{Kikukawa1995jun} (FM-KPFM) that enables the direct spatial imaging of electric field components near a surface along multiple directions simultaneously.
The lateral electric field in an FM-KPFM measurement has to date been acquired by numerically differentiating the measured surface potential versus position data to obtain the electric field versus position \cite{Burgi2002apr,Silveira2004sep,Ng2007feb,Slinker2007nov,Pingree2007dec}.
Here we show how to microscopically measure multiple electric field components simultaneously using sample-position modulation and lock-in detection. 
In measurements of the electric field perpendicular to the fast-scan direction, our position-modulation technique improved spatial resolution by a factor of four compared to numerical differentiation of the FM-KPFM surface potential image.

Microscopically measuring electric fields can be helpful for understanding both device physics and materials properties.
B\"{u}rgi \latin{et al.}\ showed experimentally that the potential measured by FM-KPFM above a transistor reflected the electrostatic potential of the accumulation layer at the transistor's buried semiconductor-insulator interface \cite{Burgi2002apr}.
This finding was subsequently justified theoretically by Silveira, Dunlap, and coworkers \cite{Silveira2007}.
Building on this observation, B\"{u}rgi and coworkers introduced the idea of using the locally-inferred electric field, the locally-inferred electrostatic potential, and the measured bulk current to infer the mobility at each location in the channel of a polymer field-effect transistor \cite{Burgi2002apr}.
The thus-measured mobility was analyzed as a function of temperature and local electric field to draw conclusions concerning charge-transport mechanisms in the polymeric semiconducting material.
If an abrupt voltage drop is apparent at a transistor contact, then the contact resistance can be computed by dividing the observed voltage drop by the measured current and likewise studied versus temperature and injecting-contact composition \cite{Burgi2003nov}.  
In samples where no such voltage drop is apparent, Silveira et al.\ showed that charge injection could nevertheless be studied microscopically by simultaneously measuring the device current and the lateral electric field at the injecting contact as a function of applied voltage and temperature; plots of the current versus the electric field could be directly compared to charge-injection theory \cite{Silveira2004sep,Ng2007feb}.
This procedure was used to assign an ``ohmicity'' to the metal-organic contact in a two-terminal device exhibiting no potential drop at the injecting contact\cite{Ng2006aug}.
Recent work has extended FM-KPFM's ability to map the distribution of trapped or mobile charges \cite{Ando2014nov,Murawski2015oct,Murawski2015dec,Yamagishi2016feb} as a function of time or frequency.
Experimental protocols have also been developed to allow FM-KPFM to make quantitative measurements of surface potential even in the presence of parasitic capacitances \cite{Ito2011may,Fuller2013feb}.
The method introduced here was designed to expand KPFM's ability to make electric field measurements with high spatial resolution.


\begin{figure*}
\includegraphics{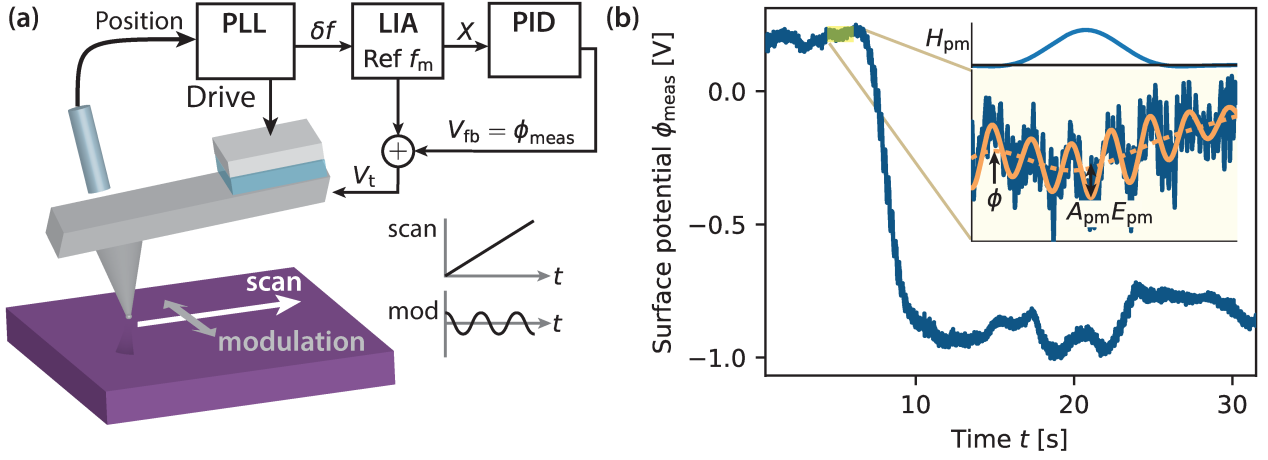}
\caption{Experimental setup and data processing.
(a) The cantilever position signal was filtered and phase-shifted by the phase-locked loop (PLL). The PLL drove the cantilever at its resonance frequency and measured the cantilever frequency shift $\delta f$.
A lock-in amplifier (LIA) measured the component of $\delta f$ at the voltage-modulation frequency $f\st{m}$.
A proportional-integral-derivative (PID) controller nulled the LIA $X$-channel output by adjusting the feedback voltage $V\st{fb}$.
Below, the sample was simultaneously scanned using a linear ramp pattern and modulated using a sinusoidal waveform at the position-modulation frequency.
(b) The surface potential-\latin{versus}-time data measures the surface potential and electric field in the scan direction at low frequencies and the electric field in the modulation direction at the modulation frequency.
Inset: Using the filter $H\st{pm}$ (top), the raw data (bottom, dark curve) was processed using a software lock-in amplifier to determine the scan-direction surface potential (light dashed curve), and modulation-direction electric field (light sinusoid).
Experimental parameters: position-modulation amplitude $A\st{pm} = \SI{45}{\nm}$, frequency $f\st{pm} = \SI{4.5}{\Hz}$, tip velocity $v=\SI{414}{\nm\per\s}$, tip-sample separation $h = \SI{200}{\nm}$, cantilever zero-to-peak amplitude $A = \SI{50}{\nm}$.
}
\label{fig:expt-diagram}
\end{figure*}

Below we report measurements over an organic field-effect transistor made from the hole-transporting small molecule DPh-BTBT \cite{Takimiya2006oct}.
We used DPh-BTBT because it is an air-stable small molecule that can be easily evaporated to produce high-mobility field-effect transistors ($\mu = \SI{2}{\cm\squared\per\V\per\s}$).
To fabricate the transistor, we evaporated \num{100} nanometers of DPh-BTBT onto a room-temperature transistor substrate at a rate of $\sim \SI{1}{\nm\per\s}$.
The transistor substrates were cleaned before use by sonicating in 1:1 acetone:isopropyl alcohol for 15~min, scrubbing and sonicating 10~min with distilled water and detergent (Aquet), sonicating in distilled water 10~min, and ozone cleaning for 5~min.
The transistor substrate was comprised of a highly $n$-doped silicon gate, a \SI{300}{\nm} thermally grown silicon oxide insulator layer, and \SI{40}{\nm}-thick gold  source and drain electrodes with a \SI{5}{\nm} chromium adhesion layer.
The electrodes were deposited using thermal evaporation and patterned into an interdigitated array.
The channel length was \SI{5}{\um} and the total channel width was \SI{19.8}{\cm}.

In FM-KPFM (Fig.~\ref{fig:expt-diagram}a), the sample's surface potential is determined by oscillating the cantilever at its resonance frequency using a phase-locked loop controller and nulling the cantilever frequency shift $\delta f$ induced by tip-sample electrostatic forces
\begin{equation}
\delta f = -\frac{f\st{c}}{4k\st{c}} C'' (V\st{t} - \phi)^2,
\end{equation}
with $f\st{c}$ the cantilever resonance frequency, $k\st{c}$ the cantilever spring constant, $C''$ the second derivative of the tip-sample capacitance with respect to the vertical direction, $V\st{t}$ the cantilever tip voltage, and $\phi$ the sample's surface potential.
The applied tip voltage is the sum of a fixed-frequency modulation voltage and a feedback voltage $V\st{fb}$: $V\st{t} = V\st{m} \sin ( 2 \pi f\st{m} t) + V\st{fb}$, where $V\st{m}$ is the voltage-modulation amplitude and $f\st{m}$ is the voltage-modulation frequency.
A lock-in amplifier measures the oscillating frequency shift at the modulation frequency
\begin{equation}
\delta f(f\st{m}) = -\frac{f\st{c}}{2k\st{c}} C'' V\st{m} (V\st{fb} - \phi).
\end{equation}
A proportional-integral-derivative controller feedback loop adjusts $V\st{fb}$ to maintain $\delta f(f\st{m})$ at zero.
With large enough feedback gain, $\delta f(f\st{m}) \approx 0$, and the feedback voltage tracks the surface potential closely: $V\st{fb} \approx \phi$.
The feedback voltage is the measured surface potential.
The assumption that $V\st{fb} = \phi$ is only valid at low frequencies or long times.
The feedback voltage $V\st{fb}$ also varies due to the effects of detector noise, low-frequency position noise, and surface potential fluctuations.

Many KPFM measurements derive information mainly from contrast in surface potential images or the average difference in surface potential over different regions of the sample.
These properties are relatively insensitive to feedback loop dynamics, noise, and surface potential fluctuations.
In contrast, these sources of error affect the calculated electric field dramatically.
To highlight the effect of these error sources, we write the measured surface potential as
\begin{equation}
\phi\st{meas} = V\st{fb} = H \ast (\phi + \phi\st{n})
\label{eq:phi_meas}
\end{equation}
where $H$ is the feedback loop's impulse response function, $\ast$ denotes convolution in the time domain, $\phi$ is the sample's actual surface potential and $\phi\st{n}$ is an equivalent surface potential noise that accounts for noise in $\phi\st{meas}$.

Noise in the measured surface potential typically arises from two sources: cantilever position noise and low-frequency surface potential noise.
The effect of cantilever position noise can be minimized by operating at a sufficiently large modulation voltage or cantilever amplitude.
Low-frequency surface potential noise could arise from position hysteresis and noise or real surface potential fluctuations caused by trapped charge or dielectric fluctuations \cite{Hoepker2011oct,Lekkala2012sep,Lekkala2013nov}.
In either case, the effect of surface potential noise can be mitigated by increasing the scan speed.

Increasing the scan speed, however, comes at a cost.
The feedback loop response function $H$ has a bandwidth $b$.
This temporal bandwidth limits the spatial resolution of the measured surface potential and electric field when the tip is scanned \cite{Garrett2016jun}.
For a tip velocity $v$, the measurement response function distorts the surface potential and electric field when they change on a length scale smaller than $x\st{res} = v/(2\pi b)$.
We find significant distortion near the contact of a DPh-BTBT transistor when $x\st{res} > \SI{10}{\nm}$ (Fig.~S2).

If the scan speed is carefully optimized, low-frequency surface potential noise along the scan axis can be avoided without distorting the measured electric field significantly.
In a 2D raster scan, however, the electric field measured along the slow scan axis will still be subject to large low-frequency ($<\SI{1}{\Hz}$) surface potential noise caused by position hysteresis and slow surface potential fluctuations.


To avoid this low-frequency noise, we modify the KPFM measurement by adding a small position modulation $\vec{\delta r}$.
The position modulation allows us to measure the electric-field component along the position-modulation direction $E\st{pm}$, since to first order in $\vec{\delta r}$
\begin{equation}
\phi(\vec{r} + \vec{\delta r}(t)) \approx \phi(\vec{r}) + \nabla \phi \cdot \vec{\delta r} = \phi(r) - \vec{E} \cdot \vec{\delta r}(t) .
\label{eq:phi-pm-FM-KPFM}
\end{equation}
We use a modulation having a direction $\hat{\delta}r = \hat{x}$ or $\hat{y}$ and a magnitude
\begin{equation}
\delta r(t) = A\st{pm} \sin (2 \pi f\st{pm} t) ,
\end{equation}
with $A\st{pm}$ the modulation amplitude and $f\st{pm}$ the modulation frequency.
We detect the electric field as an oscillating potential at the modulation frequency with amplitude $\delta \phi (f\st{pm}) = A\st{pm} E\st{pm}$.
To measure $E\st{pm}$ accurately, the modulation amplitude $A\st{pm}$ must be small enough that the potential can be approximated to first order in $\delta r$ as in Eq.~\ref{eq:phi-pm-FM-KPFM}.
The position-modulation technique could be combined with any KPFM technique that may be used to measure the sample's surface potential, including amplitude-modulation KPFM \cite{Nonnenmacher1991jun}, heterodyne KPFM \cite{Sugawara2012may,Garrett2016jun}, dissipative KPFM \cite{Miyahara2015nov,Miyahara2017apr}, or open-loop KPFM \cite{Takeuchi2007aug,Collins2013nov}.
We demonstrate the position-modulation technique in combination with FM-KPFM in this paper and call the combined protocol pm-FM-KPFM.
Because we detect $\delta \phi(f\st{pm})$ using the FM-KPFM feedback loop, the  modulation frequency $f\st{pm}$ must be significantly smaller than the feedback loop bandwidth $b$.
In our measurements, we used $A\st{pm} = \SI{30}{\nm}$ and $f\st{pm} = \SI{4.5}{\Hz}$, with $b = \SI{29}{\Hz}$ over the gate and $b=\SI{34}{\Hz}$ over the source/drain electrodes (noting that $b \propto C''$).

To perform the pm-FM-KPFM measurement, we used the experimental setup from Figure~\ref{fig:expt-diagram}a and saved the measured surface potential $\phi\st{meas}$-\latin{versus}-time data (digitized at \SI{8.192}{\kHz}).
The surface potential-\latin{versus}-time data measures the surface potential and electric field in the scan direction at low frequencies and the electric field in the modulation direction at the modulation frequency.
We low-pass filtered $\phi\st{meas}$ using the filter $H\st{pm}$ to estimate the sample surface potential along the scan direction.
We processed $\phi\st{meas}$ again using a software lock-in amplifier with lock-in filter $H\st{pm}$ to extract the electric field along the modulation direction
\begin{equation}
 E\st{pm} = \frac{\delta \phi(f\st{pm})}{A\st{pm}}
\label{eq:Epm}
\end{equation}
where $\delta \phi(f\st{pm})$ is the in-phase component of the software lock-in amplifier (supporting material~S3).
In writing Eq.~\ref{eq:Epm}, we neglect a correction term that depends on the spatial dependence of the tip-sample capacitance $C''$, the sample topography, and the position-modulation amplitude $A\st{pm}$.
For our sample and experimental conditions, the correction term would cause a worst-case fractional error in the electric field $E\st{pm}$ of less than 0.1 percent (supporting material~S4).
If this correction term were problematically large, changes in $C''$ could be corrected for by using the component of the cantilever frequency shift at $2f\st{m}$, as in open-loop KPFM measurements \cite{Takeuchi2007aug,Collins2013nov}.
The Fig.~\ref{fig:expt-diagram}b inset shows this analysis in a representative region near the contact where $\phi$ along the scan direction is relatively constant and the electric field along the modulation direction is significant.

\begin{figure}
\includegraphics{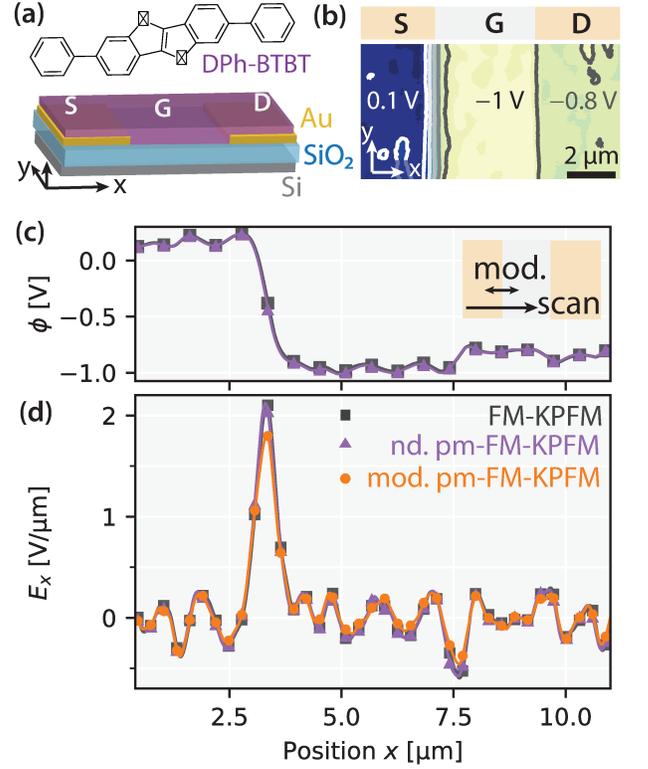}
\caption{
Demonstration of pm-FM-KPFM on a transistor.
(a) Cartoon of the transistor. 
(b) FM-KPFM image of the transistor channel, acquired with tip-sample separation $h = \SI{150}{\nm}$, zero-to-peak oscillation amplitude $A = \SI{50}{\nm}$, transistor source, gate, and drain voltages $V\st{S} = \SI{0}{\V}$, $V\st{G} = \SI{-10}{\V}$, and $V\st{D}=\SI{-1}{\V}$ respectively.
Lines show contours at \num{-0.9}, \num{-0.7}, \num{-0.4}, \num{-0.1}, and \SI{0.1}{\V}.
(c)
Surface potential measured using FM-KPFM (squares) and pm-FM-KPFM (triangles).
The scan and modulation are both along the $x$-axis (inset).
(d)
Comparison of the electric fields measured using FM-KPFM and pm-FM-KPFM. Electric field calculated by taking the numerical derivative of the KPFM surface potential (squares), the numerical derivative of the pm-FM-KPFM surface potential (triangles), and filtering the modulation component of the pm-FM-KPFM surface potential (circles).
Experimental parameters: $A\st{pm} = \SI{30}{\nm}$, $f\st{pm} = \SI{4.5}{\Hz}$, $v=\SI{0.37}{\um\per\s}$, voltage-modulation amplitude $V\st{m} = \SI{2}{\V}$ and frequency $f\st{m} = \SI{160}{\Hz}$.
}
\label{fig:pm-kpfm-1}
\end{figure}


To verify the accuracy of the electric field calculated using pm-FM-KPFM, we performed pm-FM-KPFM and FM-KPFM line scans across a DPh-BTBT thin-film transistor (Fig.~\ref{fig:pm-kpfm-1}).
So that both techniques measure $E\st{pm}$, we applied the position modulation along the scan axis (Fig.~\ref{fig:pm-kpfm-1}c inset).
The data in Fig.~\ref{fig:pm-kpfm-1}c confirms that the two techniques measure the same surface potential $\phi$.
We low-pass filtered the surface potential at $\SI{0.8}{\Hz}$, which  corresponds to a spatial frequency low-pass filter at $\nu = \SI{2.2}{\per\um}$.

From the two line scans, we calculated the electric field $E\st{pm}$ three ways (Fig.~\ref{fig:pm-kpfm-1}d).
We numerically differentiated the FM-KPFM surface potential (squares) and the pm-FM-KPFM surface potential (triangles).
To calculate the electric field from the position-modulation signal, we processed the raw surface potential data using a software lock-in amplifier whose reference frequency was set equal to the position-modulation frequency $f\st{pm} = \SI{4.5}{\Hz}$.
To make a fair comparison to standard FM-KPFM, we used a $\SI{0.8}{\Hz}$ bandwidth lock-in amplifier filter, identical to the filter used for the surface potential.
We plot the electric field $E\st{pm} = X\st{LI}/A\st{pm}$, where $X\st{LI}$ is the in-phase channel of the phased lock-in amplifier output (circles).
The electric field and surface potential calculated from pm-FM-KPFM agree with the FM-KPFM electric field and surface potential.
At equivalent bandwidth, all three electric field measurements have similar noise.

\begin{figure}
\includegraphics{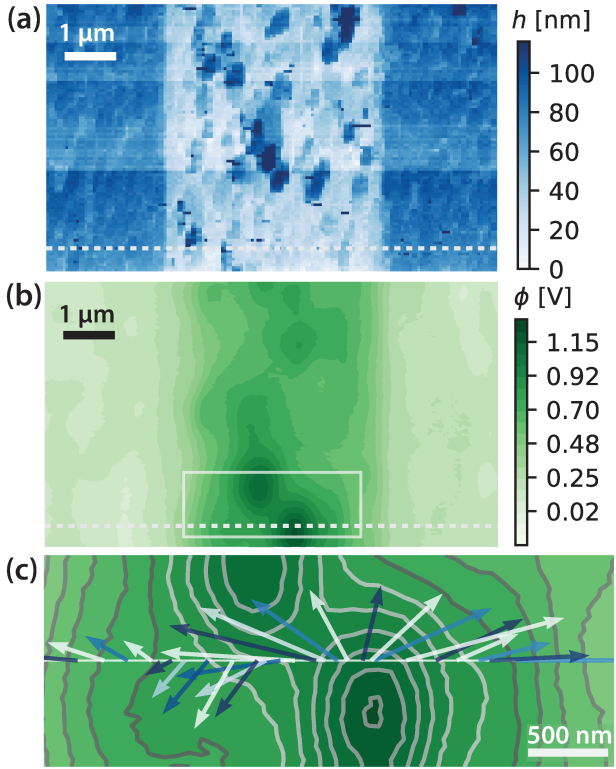}
\caption{Demonstration of pm-FM-KPFM vector electrometer.
(a) AFM image of height $h$ over the transistor channel. Full scale is \SI{120}{\nm}. The image was clipped at the data's 99th percentile for clarity.
(b) FM-KPFM image of surface potential $\phi$ over the transistor channel with $V\st{S}=V\st{D}=V\st{G}=0$. Contours are shown every \SI{75}{\milli\V}.
(c) An expanded view of the FM-KPFM surface contours in the boxed region of (b), with the vector electric field calculated from a pm-FM-KPFM linescan. The electric field vectors are colored differently for clarity.  
KPFM experimental parameters: tip-sample separation $h = \SI{200}{\nm}$, zero-to-peak amplitude $A = \SI{50}{\nm}$, tip velocity $v=\SI{1.55}{\um\per\s}$, scan spacing along the slow scan axis $\Delta y = \SI{50}{\nm}$, scan spacing along the fast scan axis $\Delta x = \SI{90}{\nm}$.
pm-FM-KPFM experimental parameters:  $A\st{pm} = \SI{30}{\nm}$, $f\st{pm} = \SI{4.5}{\Hz}$, $v=\SI{0.37}{\um\per\s}$.
}
\label{fig:pm-kpfm-2}
\end{figure}

\begin{figure}
\includegraphics{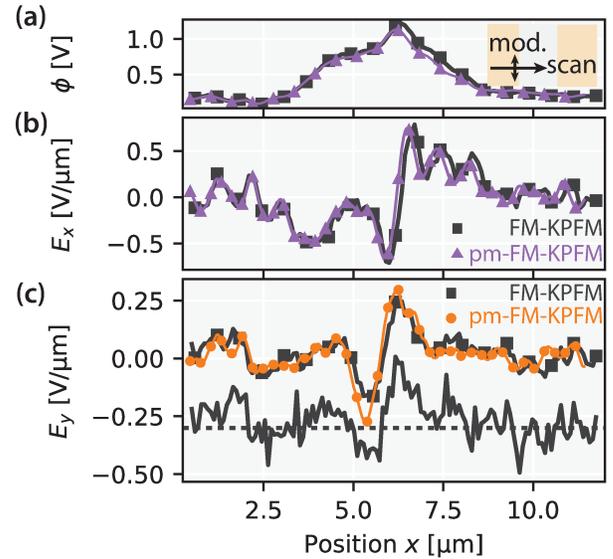}
\caption{
(a)
Surface potential measured using FM-KPFM (squares) and pm-FM-KPFM (triangles).
The scan was along the $x$-axis and the modulation was along the $y$-axis (inset).
(b)
Electric fields measured along the scan axis.
Numerical derivatives of the FM-KPFM surface potential (squares) and the pm-FM-KPFM surface potential (triangles).
(c)
Electric fields measured along the slow-scan, modulation axis $E_y$ by pm-FM-KPFM (light circles) and FM-KPFM.
The upper line (squares) show $E_y$ determined after filtering the FM-KPFM-derived surface potential along the $y$-axis (bandwidth $\nu_y^{\text{3-dB}} = \SI{2.2}{\per\um}$, see supporting material~S5).
The lower, vertically offset line shows the unfiltered FM-KPFM $E_y$ (bandwidth $\nu_y^{\text{3-dB}} = \SI{4.4}{\per\um}$).
Experimental parameters are given in Fig.~\ref{fig:pm-kpfm-2}.
}
\label{fig:pm-kpfm-3}
\end{figure}

Above we argued that pm-FM-KPFM should be useful to measure the electric field along the slow scan axis with greater signal-to-noise.
As a demonstration, we collected AFM and FM-KPFM images over the DPh-BTBT transistor with source, gate, and drain voltages set to zero (Fig.~\ref{fig:pm-kpfm-2}(a,b)).
The FM-KPFM image revealed pockets of trapped charge in the transistor channel (dark spots in box in Fig.~\ref{fig:pm-kpfm-2}b).

To probe the electric field near these trapped charges, we took a pm-FM-KPFM linescan (Fig.~\ref{fig:pm-kpfm-2}c; Fig.~\ref{fig:pm-kpfm-3}).
We applied the position modulation perpendicular to the scan direction so that we measured $E_x$ and $E_y$ simultaneously;
$E_x$ was determined by numerically differentiating the measured $\phi$ with respect to the fast scan direction while $E_y$ was obtained from $X\st{LI}$ as discussed above.
Figure~\ref{fig:pm-kpfm-2}c shows the KPFM image contours along with the in-plane electric field vector $(E_x, E_y)$ measured by pm-FM-KPFM.
One consequence of the electric field being the negative gradient of the electrostatic potential is that the electric field vector at location $\vec{r}$ must be perpendicular to a line tangent to the constant-$\phi(x,y)$ surface passing through $\vec{r}$.
This perpendicular relationship is clearly evident in Fig.~\ref{fig:pm-kpfm-2}c, demonstrating pm-FM-KPFM's ability to serve as a \emph{vector electrometer}.

In Figure~\ref{fig:pm-kpfm-3}, we quantitatively compare the surface potential and the electric field measured by FM-KPFM and pm-FM-KPFM.
Both measurements computed the electric field along the fast scan axis ($E_x$) by numerical differentiation, using surface potential data averaged for \SI{200}{\ms} in each case.
The two measurements of $E_x$ are in close agreement (Fig.~\ref{fig:pm-kpfm-3}b).
We evaluate spatial resolution using the spatial frequency 3-dB bandwidth, defined as the frequency at which the measured electric field captures $70.8$ percent of the actual electric field signal.
The measurements of $E_x$ both have a bandwidth of $\nu_x^{\text{3-dB}} = \SI{2.2}{\per\um}$.

Figure~\ref{fig:pm-kpfm-3}c shows the electric field along the slow scan axis $E_y$.
We plot the pm-FM-KPFM $E_y$ (light circles) along with two calculations of $E_y$ obtained from the FM-KPFM image of the sample's surface potential.
The dark squares show a filtered FM-KPFM $E_y$, with the 3-dB spatial bandwidth along the $y$-axis limited to $\nu_y^{\text{3-dB}} = \SI{2.2}{\per\um}$ using a 17-point low-pass filter.
The pm-FM-KPFM spatial bandwidth is $\nu_y^{\text{3-dB}} = \SI{8.6}{\per\um}$, limited by the magnitude of the position-modulation amplitude $A\st{pm}$.
Even with a bandwidth along the $y$-axis nearly 4-times greater, the pm-FM-KPFM electric field has similar or lower noise than the FM-KPFM electric field.

For comparison, in Fig.~\ref{fig:pm-kpfm-3}c we also we plot the unfiltered FM-KPFM $E_y$ (offset by \SI{-0.3}{\V\per\um}).
The unfiltered $E_y$ has a spatial bandwidth $\nu_y^{\text{3-dB}} = \SI{4.4}{\per\um}$, limited by the spacing between data points $\Delta y = \SI{50}{\nm}$.
Despite having a factor of two lower bandwidth than the pm-FM-KPFM measurement, the FM-KPFM signal exhibits significantly worse noise.
We can understand this observation by noting that the FM-KPFM $E_y$ was computed by subtracting surface potential points acquired \SI{8}{\s} apart.
Slow drift in the surface potential on this timescale thus shows up as noise in the FM-KPFM-inferred $E_y$ \cite{Schumacher2016apr}.
Viewing the measurements in the frequency domain, the FM-KPFM measurement of $E_y$ incorporates surface potential noise at temporal frequencies near $(\SI{8}{\s}/\text{line})^{-1} \sim \SI{0.1}{\Hz}$.
The pm-FM-KPFM measurement of $E_y$ incorporates surface potential noise at frequencies near $f\st{pm} = \SI{4.5}{\Hz}$, where overall surface potential noise is near a minimum (Fig.~S6).
This advantage in signal-to-noise ratio can be used to achieve higher spatial resolution at equivalent bandwidth or lower noise at equivalent spatial resolution.
Moreover, pm-FM-KPFM allows $E_y$ to be measured simultaneously with $E_x$.

We anticipate that the simple modification of KPFM introduced here will facilitate electric field measurements in a variety of systems.
It is increasingly recognized that transistor measurements significantly overestimate charge mobility in high-performance organic semiconductors\cite{McCulloch2016jun,Bittle2016mar}; 1D electric field mapping, in conjunction with current measurements, offers a general route to avoiding this materials-characterization pitfall.
Although we demonstrate its use for measuring lateral electric fields, it should also be possible to measure vertical electric fields with an additional vertical position modulation.
The local electric field vectors measured here are already an advance from lateral electric field line scans, and we envision applying the pm-FM-KPFM technique to measure local electric fields in bulk heterojunction solar cell blends \cite{Luria2012nov,Tennyson2015sep} and other composite materials \cite{Cadena2016apr}.
KPFM measurements mimic device operation near the  open-circuit voltage $V\st{OC}$ condition; acquiring 2D electric field scans would allow the visualization of the current flow direction at domain boundaries in illuminated films near the $V\st{OC}$ condition.


The supplementary materials contain details of the experimental setup, data analysis, and experimental noise.
The experimental data are available online \cite{Dwyer2017jul_data}.
The authors acknowledge support from Cornell University and the National Science Foundation through an NSF Graduate Research Fellowship (L.M.S.), NSF-DMR 1309540, and NSF-DMR 1602951.
This work was performed in part at the Cornell NanoScale Facility, a member of the National Nanotechnology Coordinated Infrastructure (NNCI), which is supported by the National Science Foundation (Grant ECCS-1542081).

\def\bibsection{\vspace{6pt}}
\setlength{\bibsep}{0pt}

\bibliography{Dwyer}

\begin{thebibliography}{33}%
\makeatletter
\providecommand \@ifxundefined [1]{%
 \@ifx{#1\undefined}
}%
\providecommand \@ifnum [1]{%
 \ifnum #1\expandafter \@firstoftwo
 \else \expandafter \@secondoftwo
 \fi
}%
\providecommand \@ifx [1]{%
 \ifx #1\expandafter \@firstoftwo
 \else \expandafter \@secondoftwo
 \fi
}%
\providecommand \natexlab [1]{#1}%
\providecommand \enquote  [1]{``#1''}%
\providecommand \bibnamefont  [1]{#1}%
\providecommand \bibfnamefont [1]{#1}%
\providecommand \citenamefont [1]{#1}%
\providecommand \href@noop [0]{\@secondoftwo}%
\providecommand \href [0]{\begingroup \@sanitize@url \@href}%
\providecommand \@href[1]{\@@startlink{#1}\@@href}%
\providecommand \@@href[1]{\endgroup#1\@@endlink}%
\providecommand \@sanitize@url [0]{\catcode `\\12\catcode `\$12\catcode
  `\&12\catcode `\#12\catcode `\^12\catcode `\_12\catcode `\%12\relax}%
\providecommand \@@startlink[1]{}%
\providecommand \@@endlink[0]{}%
\providecommand \url  [0]{\begingroup\@sanitize@url \@url }%
\providecommand \@url [1]{\endgroup\@href {#1}{\urlprefix }}%
\providecommand \urlprefix  [0]{URL }%
\providecommand \Eprint [0]{\href }%
\providecommand \doibase [0]{http://dx.doi.org/}%
\providecommand \selectlanguage [0]{\@gobble}%
\providecommand \bibinfo  [0]{\@secondoftwo}%
\providecommand \bibfield  [0]{\@secondoftwo}%
\providecommand \translation [1]{[#1]}%
\providecommand \BibitemOpen [0]{}%
\providecommand \bibitemStop [0]{}%
\providecommand \bibitemNoStop [0]{.\EOS\space}%
\providecommand \EOS [0]{\spacefactor3000\relax}%
\providecommand \BibitemShut  [1]{\csname bibitem#1\endcsname}%
\let\auto@bib@innerbib\@empty
\bibitem [{\citenamefont {Kikukawa}, \citenamefont {Hosaka},\ and\
  \citenamefont {Imura}(1995)}]{Kikukawa1995jun}%
  \BibitemOpen
  \bibfield  {author} {\bibinfo {author} {\bibfnamefont {A.}~\bibnamefont
  {Kikukawa}}, \bibinfo {author} {\bibfnamefont {S.}~\bibnamefont {Hosaka}}, \
  and\ \bibinfo {author} {\bibfnamefont {R.}~\bibnamefont {Imura}},\ }\href
  {\doibase 10.1063/1.113780} {\bibfield  {journal} {\bibinfo  {journal} {Appl.
  Phys. Lett.}\ }\textbf {\bibinfo {volume} {66}},\ \bibinfo {pages} {3510 }
  (\bibinfo {year} {1995})}\BibitemShut {NoStop}%
\bibitem [{\citenamefont {Burgi}, \citenamefont {Sirringhaus},\ and\
  \citenamefont {Friend}(2002)}]{Burgi2002apr}%
  \BibitemOpen
  \bibfield  {author} {\bibinfo {author} {\bibfnamefont {L.}~\bibnamefont
  {Burgi}}, \bibinfo {author} {\bibfnamefont {H.}~\bibnamefont {Sirringhaus}},
  \ and\ \bibinfo {author} {\bibfnamefont {R.}~\bibnamefont {Friend}},\ }\href
  {\doibase 10.1063/1.1470702} {\bibfield  {journal} {\bibinfo  {journal}
  {Appl. Phys. Lett.}\ }\textbf {\bibinfo {volume} {80}},\ \bibinfo {pages}
  {2913 } (\bibinfo {year} {2002})}\BibitemShut {NoStop}%
\bibitem [{\citenamefont {Silveira}\ and\ \citenamefont
  {Marohn}(2004)}]{Silveira2004sep}%
  \BibitemOpen
  \bibfield  {author} {\bibinfo {author} {\bibfnamefont {W.~R.}\ \bibnamefont
  {Silveira}}\ and\ \bibinfo {author} {\bibfnamefont {J.~A.}\ \bibnamefont
  {Marohn}},\ }\href {\doibase 10.1103/PhysRevLett.93.116104} {\bibfield
  {journal} {\bibinfo  {journal} {Phys. Rev. Lett.}\ }\textbf {\bibinfo
  {volume} {93}},\ \bibinfo {pages} {116104} (\bibinfo {year}
  {2004})}\BibitemShut {NoStop}%
\bibitem [{\citenamefont {Ng}, \citenamefont {Silveira},\ and\ \citenamefont
  {Marohn}(2007)}]{Ng2007feb}%
  \BibitemOpen
  \bibfield  {author} {\bibinfo {author} {\bibfnamefont {T.~N.}\ \bibnamefont
  {Ng}}, \bibinfo {author} {\bibfnamefont {W.~R.}\ \bibnamefont {Silveira}}, \
  and\ \bibinfo {author} {\bibfnamefont {J.~A.}\ \bibnamefont {Marohn}},\
  }\href {\doibase 10.1103/PhysRevLett.98.066101} {\bibfield  {journal}
  {\bibinfo  {journal} {Phys. Rev. Lett.}\ }\textbf {\bibinfo {volume} {98}},\
  \bibinfo {pages} {066101} (\bibinfo {year} {2007})}\BibitemShut {NoStop}%
\bibitem [{\citenamefont {Slinker}\ \emph {et~al.}(2007)\citenamefont
  {Slinker}, \citenamefont {DeFranco}, \citenamefont {Jaquith}, \citenamefont
  {Silveira}, \citenamefont {Zhong}, \citenamefont {Moran-Mirabal},
  \citenamefont {Craighead}, \citenamefont {Abru\~{n}a}, \citenamefont
  {Marohn},\ and\ \citenamefont {Malliaras}}]{Slinker2007nov}%
  \BibitemOpen
  \bibfield  {author} {\bibinfo {author} {\bibfnamefont {J.~D.}\ \bibnamefont
  {Slinker}}, \bibinfo {author} {\bibfnamefont {J.~A.}\ \bibnamefont
  {DeFranco}}, \bibinfo {author} {\bibfnamefont {M.~J.}\ \bibnamefont
  {Jaquith}}, \bibinfo {author} {\bibfnamefont {W.~R.}\ \bibnamefont
  {Silveira}}, \bibinfo {author} {\bibfnamefont {Y.-W.}\ \bibnamefont {Zhong}},
  \bibinfo {author} {\bibfnamefont {J.~M.}\ \bibnamefont {Moran-Mirabal}},
  \bibinfo {author} {\bibfnamefont {H.~G.}\ \bibnamefont {Craighead}}, \bibinfo
  {author} {\bibfnamefont {H.~D.}\ \bibnamefont {Abru\~{n}a}}, \bibinfo
  {author} {\bibfnamefont {J.~A.}\ \bibnamefont {Marohn}}, \ and\ \bibinfo
  {author} {\bibfnamefont {G.~G.}\ \bibnamefont {Malliaras}},\ }\href {\doibase
  10.1038/nmat2021} {\bibfield  {journal} {\bibinfo  {journal} {Nat. Mater.}\
  }\textbf {\bibinfo {volume} {6}},\ \bibinfo {pages} {894 } (\bibinfo {year}
  {2007})}\BibitemShut {NoStop}%
\bibitem [{\citenamefont {Pingree}\ \emph {et~al.}(2007)\citenamefont
  {Pingree}, \citenamefont {Rodovsky}, \citenamefont {Coffey}, \citenamefont
  {Bartholomew},\ and\ \citenamefont {Ginger}}]{Pingree2007dec}%
  \BibitemOpen
  \bibfield  {author} {\bibinfo {author} {\bibfnamefont {L.~S.~C.}\
  \bibnamefont {Pingree}}, \bibinfo {author} {\bibfnamefont {D.~B.}\
  \bibnamefont {Rodovsky}}, \bibinfo {author} {\bibfnamefont {D.~C.}\
  \bibnamefont {Coffey}}, \bibinfo {author} {\bibfnamefont {G.~P.}\
  \bibnamefont {Bartholomew}}, \ and\ \bibinfo {author} {\bibfnamefont {D.~S.}\
  \bibnamefont {Ginger}},\ }\href {\doibase 10.1021/ja074760m} {\bibfield
  {journal} {\bibinfo  {journal} {J. Am. Chem. Soc.}\ }\textbf {\bibinfo
  {volume} {129}},\ \bibinfo {pages} {15903} (\bibinfo {year} {2007})},\
  \bibinfo {note} {pMID: 18052165}\BibitemShut {NoStop}%
\bibitem [{\citenamefont {Silveira}\ \emph {et~al.}(2007)\citenamefont
  {Silveira}, \citenamefont {Muller}, \citenamefont {Ng}, \citenamefont
  {Dunlap},\ and\ \citenamefont {Marohn}}]{Silveira2007}%
  \BibitemOpen
  \bibfield  {author} {\bibinfo {author} {\bibfnamefont {W.~R.}\ \bibnamefont
  {Silveira}}, \bibinfo {author} {\bibfnamefont {E.~M.}\ \bibnamefont
  {Muller}}, \bibinfo {author} {\bibfnamefont {T.~N.}\ \bibnamefont {Ng}},
  \bibinfo {author} {\bibfnamefont {D.~H.}\ \bibnamefont {Dunlap}}, \ and\
  \bibinfo {author} {\bibfnamefont {J.~A.}\ \bibnamefont {Marohn}},\ }in\ \href
  {http://www.amazon.com/Scanning-Probe-Microscopy-Electrical-
  Electromechanical/dp/0387286675/sr=1-1/qid=1171915254/
  ref=pd_bbs_sr_1/002-1394425-2104019?ie=UTF8&s=books} {\emph {\bibinfo
  {booktitle} {Scanning Probe Microscopy: {E}lectrical and Electromechanical
  Phenomena at the Nanoscale}}},\ Vol.~\bibinfo {volume} {II},\ \bibinfo
  {editor} {edited by\ \bibinfo {editor} {\bibfnamefont {S.~V.}\ \bibnamefont
  {Kalinin}}\ and\ \bibinfo {editor} {\bibfnamefont {A.}~\bibnamefont
  {Gruverman}}}\ (\bibinfo  {publisher} {Springer Verlag},\ \bibinfo {address}
  {New York},\ \bibinfo {year} {2007})\ pp.\ \bibinfo {pages} {788 --
  830}\BibitemShut {NoStop}%
\bibitem [{\citenamefont {Burgi}\ \emph {et~al.}(2003)\citenamefont {Burgi},
  \citenamefont {Richards}, \citenamefont {Friend},\ and\ \citenamefont
  {Sirringhaus}}]{Burgi2003nov}%
  \BibitemOpen
  \bibfield  {author} {\bibinfo {author} {\bibfnamefont {L.}~\bibnamefont
  {Burgi}}, \bibinfo {author} {\bibfnamefont {T.}~\bibnamefont {Richards}},
  \bibinfo {author} {\bibfnamefont {R.}~\bibnamefont {Friend}}, \ and\ \bibinfo
  {author} {\bibfnamefont {H.}~\bibnamefont {Sirringhaus}},\ }\href {\doibase
  10.1063/1.1613369} {\bibfield  {journal} {\bibinfo  {journal} {J. Appl.
  Phys.}\ }\textbf {\bibinfo {volume} {94}},\ \bibinfo {pages} {6129 }
  (\bibinfo {year} {2003})}\BibitemShut {NoStop}%
\bibitem [{\citenamefont {Ng}, \citenamefont {Silveira},\ and\ \citenamefont
  {Marohn}(2006)}]{Ng2006aug}%
  \BibitemOpen
  \bibfield  {author} {\bibinfo {author} {\bibfnamefont {T.~N.}\ \bibnamefont
  {Ng}}, \bibinfo {author} {\bibfnamefont {W.~R.}\ \bibnamefont {Silveira}}, \
  and\ \bibinfo {author} {\bibfnamefont {J.~A.}\ \bibnamefont {Marohn}},\
  }\href {\doibase 10.1117/12.681010} {\bibfield  {journal} {\bibinfo
  {journal} {Proc. {SPIE}}\ }\textbf {\bibinfo {volume} {6336}},\ \bibinfo
  {pages} {63360A} (\bibinfo {year} {2006})}\BibitemShut {NoStop}%
\bibitem [{\citenamefont {Ando}\ \emph {et~al.}(2014)\citenamefont {Ando},
  \citenamefont {Heike}, \citenamefont {Kawasaki},\ and\ \citenamefont
  {Hashizume}}]{Ando2014nov}%
  \BibitemOpen
  \bibfield  {author} {\bibinfo {author} {\bibfnamefont {M.}~\bibnamefont
  {Ando}}, \bibinfo {author} {\bibfnamefont {S.}~\bibnamefont {Heike}},
  \bibinfo {author} {\bibfnamefont {M.}~\bibnamefont {Kawasaki}}, \ and\
  \bibinfo {author} {\bibfnamefont {T.}~\bibnamefont {Hashizume}},\ }\href
  {\doibase 10.1063/1.4901946} {\bibfield  {journal} {\bibinfo  {journal}
  {Appl. Phys. Lett.}\ }\textbf {\bibinfo {volume} {105}},\ \bibinfo {pages}
  {193303} (\bibinfo {year} {2014})}\BibitemShut {NoStop}%
\bibitem [{\citenamefont {Murawski}\ \emph
  {et~al.}(2015{\natexlab{a}})\citenamefont {Murawski}, \citenamefont
  {Graupner}, \citenamefont {Milde}, \citenamefont {Raupach}, \citenamefont
  {Zerweck-Trogisch},\ and\ \citenamefont {Eng}}]{Murawski2015oct}%
  \BibitemOpen
  \bibfield  {author} {\bibinfo {author} {\bibfnamefont {J.}~\bibnamefont
  {Murawski}}, \bibinfo {author} {\bibfnamefont {T.}~\bibnamefont {Graupner}},
  \bibinfo {author} {\bibfnamefont {P.}~\bibnamefont {Milde}}, \bibinfo
  {author} {\bibfnamefont {R.}~\bibnamefont {Raupach}}, \bibinfo {author}
  {\bibfnamefont {U.}~\bibnamefont {Zerweck-Trogisch}}, \ and\ \bibinfo
  {author} {\bibfnamefont {L.~M.}\ \bibnamefont {Eng}},\ }\href {\doibase
  10.1063/1.4933289} {\bibfield  {journal} {\bibinfo  {journal} {J. Appl.
  Phys.}\ }\textbf {\bibinfo {volume} {118}},\ \bibinfo {pages} {154302}
  (\bibinfo {year} {2015}{\natexlab{a}})}\BibitemShut {NoStop}%
\bibitem [{\citenamefont {Murawski}\ \emph
  {et~al.}(2015{\natexlab{b}})\citenamefont {Murawski}, \citenamefont
  {M\"onch}, \citenamefont {Milde}, \citenamefont {Hein}, \citenamefont
  {Nicht}, \citenamefont {Zerweck-Trogisch},\ and\ \citenamefont
  {Eng}}]{Murawski2015dec}%
  \BibitemOpen
  \bibfield  {author} {\bibinfo {author} {\bibfnamefont {J.}~\bibnamefont
  {Murawski}}, \bibinfo {author} {\bibfnamefont {T.}~\bibnamefont {M\"onch}},
  \bibinfo {author} {\bibfnamefont {P.}~\bibnamefont {Milde}}, \bibinfo
  {author} {\bibfnamefont {M.~P.}\ \bibnamefont {Hein}}, \bibinfo {author}
  {\bibfnamefont {S.}~\bibnamefont {Nicht}}, \bibinfo {author} {\bibfnamefont
  {U.}~\bibnamefont {Zerweck-Trogisch}}, \ and\ \bibinfo {author}
  {\bibfnamefont {L.~M.}\ \bibnamefont {Eng}},\ }\href {\doibase
  10.1063/1.4938529} {\bibfield  {journal} {\bibinfo  {journal} {J. Appl.
  Phys.}\ }\textbf {\bibinfo {volume} {118}},\ \bibinfo {pages} {244502}
  (\bibinfo {year} {2015}{\natexlab{b}})}\BibitemShut {NoStop}%
\bibitem [{\citenamefont {Yamagishi}\ \emph {et~al.}(2016)\citenamefont
  {Yamagishi}, \citenamefont {Kobayashi}, \citenamefont {Noda},\ and\
  \citenamefont {Yamada}}]{Yamagishi2016feb}%
  \BibitemOpen
  \bibfield  {author} {\bibinfo {author} {\bibfnamefont {Y.}~\bibnamefont
  {Yamagishi}}, \bibinfo {author} {\bibfnamefont {K.}~\bibnamefont
  {Kobayashi}}, \bibinfo {author} {\bibfnamefont {K.}~\bibnamefont {Noda}}, \
  and\ \bibinfo {author} {\bibfnamefont {H.}~\bibnamefont {Yamada}},\ }\href
  {\doibase 10.1063/1.4943140} {\bibfield  {journal} {\bibinfo  {journal}
  {Appl. Phys. Lett.}\ }\textbf {\bibinfo {volume} {108}},\ \bibinfo {pages}
  {093302} (\bibinfo {year} {2016})}\BibitemShut {NoStop}%
\bibitem [{\citenamefont {Ito}\ \emph {et~al.}(2011)\citenamefont {Ito},
  \citenamefont {Hosokawa}, \citenamefont {Nishi}, \citenamefont {Miyato},
  \citenamefont {Kobayashi}, \citenamefont {Matsushige},\ and\ \citenamefont
  {Yamada}}]{Ito2011may}%
  \BibitemOpen
  \bibfield  {author} {\bibinfo {author} {\bibfnamefont {M.}~\bibnamefont
  {Ito}}, \bibinfo {author} {\bibfnamefont {Y.}~\bibnamefont {Hosokawa}},
  \bibinfo {author} {\bibfnamefont {R.}~\bibnamefont {Nishi}}, \bibinfo
  {author} {\bibfnamefont {Y.}~\bibnamefont {Miyato}}, \bibinfo {author}
  {\bibfnamefont {K.}~\bibnamefont {Kobayashi}}, \bibinfo {author}
  {\bibfnamefont {K.}~\bibnamefont {Matsushige}}, \ and\ \bibinfo {author}
  {\bibfnamefont {H.}~\bibnamefont {Yamada}},\ }\href {\doibase
  10.1380/ejssnt.2011.210} {\bibfield  {journal} {\bibinfo  {journal} {e-J.
  Surf. Sci. Nanotech.}\ }\textbf {\bibinfo {volume} {9}},\ \bibinfo {pages}
  {210} (\bibinfo {year} {2011})}\BibitemShut {NoStop}%
\bibitem [{\citenamefont {Fuller}\ \emph {et~al.}(2013)\citenamefont {Fuller},
  \citenamefont {Pan}, \citenamefont {Corso}, \citenamefont {Tolga~Gul},
  \citenamefont {Gomez},\ and\ \citenamefont {Collins}}]{Fuller2013feb}%
  \BibitemOpen
  \bibfield  {author} {\bibinfo {author} {\bibfnamefont {E.~J.}\ \bibnamefont
  {Fuller}}, \bibinfo {author} {\bibfnamefont {D.}~\bibnamefont {Pan}},
  \bibinfo {author} {\bibfnamefont {B.~L.}\ \bibnamefont {Corso}}, \bibinfo
  {author} {\bibfnamefont {O.}~\bibnamefont {Tolga~Gul}}, \bibinfo {author}
  {\bibfnamefont {J.~R.}\ \bibnamefont {Gomez}}, \ and\ \bibinfo {author}
  {\bibfnamefont {P.~G.}\ \bibnamefont {Collins}},\ }\href {\doibase
  10.1063/1.4793480} {\bibfield  {journal} {\bibinfo  {journal} {Appl. Phys.
  Lett.}\ }\textbf {\bibinfo {volume} {102}},\ \bibinfo {pages} {083503}
  (\bibinfo {year} {2013})}\BibitemShut {NoStop}%
\bibitem [{\citenamefont {Takimiya}\ \emph {et~al.}(2006)\citenamefont
  {Takimiya}, \citenamefont {Ebata}, \citenamefont {Sakamoto}, \citenamefont
  {Izawa}, \citenamefont {Otsubo},\ and\ \citenamefont
  {Kunugi}}]{Takimiya2006oct}%
  \BibitemOpen
  \bibfield  {author} {\bibinfo {author} {\bibfnamefont {K.}~\bibnamefont
  {Takimiya}}, \bibinfo {author} {\bibfnamefont {H.}~\bibnamefont {Ebata}},
  \bibinfo {author} {\bibfnamefont {K.}~\bibnamefont {Sakamoto}}, \bibinfo
  {author} {\bibfnamefont {T.}~\bibnamefont {Izawa}}, \bibinfo {author}
  {\bibfnamefont {T.}~\bibnamefont {Otsubo}}, \ and\ \bibinfo {author}
  {\bibfnamefont {Y.}~\bibnamefont {Kunugi}},\ }\href {\doibase
  10.1021/ja064052l} {\bibfield  {journal} {\bibinfo  {journal} {J. Am. Chem.
  Soc.}\ }\textbf {\bibinfo {volume} {128}},\ \bibinfo {pages} {12604}
  (\bibinfo {year} {2006})}\BibitemShut {NoStop}%
\bibitem [{\citenamefont {Hoepker}\ \emph {et~al.}(2011)\citenamefont
  {Hoepker}, \citenamefont {Lekkala}, \citenamefont {Loring},\ and\
  \citenamefont {Marohn}}]{Hoepker2011oct}%
  \BibitemOpen
  \bibfield  {author} {\bibinfo {author} {\bibfnamefont {N.}~\bibnamefont
  {Hoepker}}, \bibinfo {author} {\bibfnamefont {S.}~\bibnamefont {Lekkala}},
  \bibinfo {author} {\bibfnamefont {R.~F.}\ \bibnamefont {Loring}}, \ and\
  \bibinfo {author} {\bibfnamefont {J.~A.}\ \bibnamefont {Marohn}},\ }\href
  {\doibase 10.1021/jp207387d} {\bibfield  {journal} {\bibinfo  {journal} {J.
  Phys. Chem. B}\ }\textbf {\bibinfo {volume} {115}},\ \bibinfo {pages} {14493
  } (\bibinfo {year} {2011})}\BibitemShut {NoStop}%
\bibitem [{\citenamefont {Lekkala}\ \emph {et~al.}(2012)\citenamefont
  {Lekkala}, \citenamefont {Hoepker}, \citenamefont {Marohn},\ and\
  \citenamefont {Loring}}]{Lekkala2012sep}%
  \BibitemOpen
  \bibfield  {author} {\bibinfo {author} {\bibfnamefont {S.}~\bibnamefont
  {Lekkala}}, \bibinfo {author} {\bibfnamefont {N.}~\bibnamefont {Hoepker}},
  \bibinfo {author} {\bibfnamefont {J.~A.}\ \bibnamefont {Marohn}}, \ and\
  \bibinfo {author} {\bibfnamefont {R.~F.}\ \bibnamefont {Loring}},\ }\href
  {\doibase 10.1063/1.4754602} {\bibfield  {journal} {\bibinfo  {journal} {J.
  Chem. Phys.}\ }\textbf {\bibinfo {volume} {137}},\ \bibinfo {pages} {124701}
  (\bibinfo {year} {2012})}\BibitemShut {NoStop}%
\bibitem [{\citenamefont {Lekkala}, \citenamefont {Marohn},\ and\ \citenamefont
  {Loring}(2013)}]{Lekkala2013nov}%
  \BibitemOpen
  \bibfield  {author} {\bibinfo {author} {\bibfnamefont {S.}~\bibnamefont
  {Lekkala}}, \bibinfo {author} {\bibfnamefont {J.~A.}\ \bibnamefont {Marohn}},
  \ and\ \bibinfo {author} {\bibfnamefont {R.~F.}\ \bibnamefont {Loring}},\
  }\href {\doibase 10.1063/1.4828862} {\bibfield  {journal} {\bibinfo
  {journal} {J. Chem. Phys.}\ }\textbf {\bibinfo {volume} {139}},\ \bibinfo
  {pages} {184702} (\bibinfo {year} {2013})}\BibitemShut {NoStop}%
\bibitem [{\citenamefont {Garrett}\ and\ \citenamefont
  {Munday}(2016)}]{Garrett2016jun}%
  \BibitemOpen
  \bibfield  {author} {\bibinfo {author} {\bibfnamefont {J.~L.}\ \bibnamefont
  {Garrett}}\ and\ \bibinfo {author} {\bibfnamefont {J.~N.}\ \bibnamefont
  {Munday}},\ }\href {\doibase 10.1088/0957-4484/27/24/245705} {\bibfield
  {journal} {\bibinfo  {journal} {Nanotechnology}\ }\textbf {\bibinfo {volume}
  {27}},\ \bibinfo {pages} {245705} (\bibinfo {year} {2016})}\BibitemShut
  {NoStop}%
\bibitem [{\citenamefont {Nonnenmacher}, \citenamefont {O'Boyle},\ and\
  \citenamefont {Wickramasinghe}(1991)}]{Nonnenmacher1991jun}%
  \BibitemOpen
  \bibfield  {author} {\bibinfo {author} {\bibfnamefont {M.}~\bibnamefont
  {Nonnenmacher}}, \bibinfo {author} {\bibfnamefont {M.}~\bibnamefont
  {O'Boyle}}, \ and\ \bibinfo {author} {\bibfnamefont {H.}~\bibnamefont
  {Wickramasinghe}},\ }\href {\doibase 10.1063/1.105227} {\bibfield  {journal}
  {\bibinfo  {journal} {Appl. Phys. Lett.}\ }\textbf {\bibinfo {volume} {58}},\
  \bibinfo {pages} {2921 } (\bibinfo {year} {1991})}\BibitemShut {NoStop}%
\bibitem [{\citenamefont {Sugawara}\ \emph {et~al.}(2012)\citenamefont
  {Sugawara}, \citenamefont {Kou}, \citenamefont {Ma}, \citenamefont {Kamijo},
  \citenamefont {Naitoh},\ and\ \citenamefont {Jun~Li}}]{Sugawara2012may}%
  \BibitemOpen
  \bibfield  {author} {\bibinfo {author} {\bibfnamefont {Y.}~\bibnamefont
  {Sugawara}}, \bibinfo {author} {\bibfnamefont {L.}~\bibnamefont {Kou}},
  \bibinfo {author} {\bibfnamefont {Z.}~\bibnamefont {Ma}}, \bibinfo {author}
  {\bibfnamefont {T.}~\bibnamefont {Kamijo}}, \bibinfo {author} {\bibfnamefont
  {Y.}~\bibnamefont {Naitoh}}, \ and\ \bibinfo {author} {\bibfnamefont
  {Y.}~\bibnamefont {Jun~Li}},\ }\href {\doibase 10.1063/1.4723697} {\bibfield
  {journal} {\bibinfo  {journal} {Appl. Phys. Lett.}\ }\textbf {\bibinfo
  {volume} {100}},\ \bibinfo {pages} {223104} (\bibinfo {year}
  {2012})}\BibitemShut {NoStop}%
\bibitem [{\citenamefont {Miyahara}\ \emph {et~al.}(2015)\citenamefont
  {Miyahara}, \citenamefont {Topple}, \citenamefont {Schumacher},\ and\
  \citenamefont {Grutter}}]{Miyahara2015nov}%
  \BibitemOpen
  \bibfield  {author} {\bibinfo {author} {\bibfnamefont {Y.}~\bibnamefont
  {Miyahara}}, \bibinfo {author} {\bibfnamefont {J.}~\bibnamefont {Topple}},
  \bibinfo {author} {\bibfnamefont {Z.}~\bibnamefont {Schumacher}}, \ and\
  \bibinfo {author} {\bibfnamefont {P.}~\bibnamefont {Grutter}},\ }\href
  {\doibase 10.1103/PhysRevApplied.4.054011} {\bibfield  {journal} {\bibinfo
  {journal} {Phys. Rev. Appl.}\ }\textbf {\bibinfo {volume} {4}} (\bibinfo
  {year} {2015}),\ 10.1103/PhysRevApplied.4.054011}\BibitemShut {NoStop}%
\bibitem [{\citenamefont {Miyahara}\ and\ \citenamefont
  {Grutter}(2017)}]{Miyahara2017apr}%
  \BibitemOpen
  \bibfield  {author} {\bibinfo {author} {\bibfnamefont {Y.}~\bibnamefont
  {Miyahara}}\ and\ \bibinfo {author} {\bibfnamefont {P.}~\bibnamefont
  {Grutter}},\ }\href {\doibase 10.1063/1.4981937} {\bibfield  {journal}
  {\bibinfo  {journal} {Appl. Phys. Lett.}\ }\textbf {\bibinfo {volume}
  {110}},\ \bibinfo {pages} {163103} (\bibinfo {year} {2017})}\BibitemShut
  {NoStop}%
\bibitem [{\citenamefont {Takeuchi}\ \emph {et~al.}(2007)\citenamefont
  {Takeuchi}, \citenamefont {Ohrai}, \citenamefont {Yoshida},\ and\
  \citenamefont {Shigekawa}}]{Takeuchi2007aug}%
  \BibitemOpen
  \bibfield  {author} {\bibinfo {author} {\bibfnamefont {O.}~\bibnamefont
  {Takeuchi}}, \bibinfo {author} {\bibfnamefont {Y.}~\bibnamefont {Ohrai}},
  \bibinfo {author} {\bibfnamefont {S.}~\bibnamefont {Yoshida}}, \ and\
  \bibinfo {author} {\bibfnamefont {H.}~\bibnamefont {Shigekawa}},\ }\href
  {\doibase 10.1143/JJAP.46.5626} {\bibfield  {journal} {\bibinfo  {journal}
  {Jpn. J. Appl. Phys.}\ }\textbf {\bibinfo {volume} {46}},\ \bibinfo {pages}
  {5626} (\bibinfo {year} {2007})}\BibitemShut {NoStop}%
\bibitem [{\citenamefont {Collins}\ \emph {et~al.}(2013)\citenamefont
  {Collins}, \citenamefont {Kilpatrick}, \citenamefont {Weber}, \citenamefont
  {Tselev}, \citenamefont {Vlassiouk}, \citenamefont {Ivanov}, \citenamefont
  {Jesse}, \citenamefont {Kalinin},\ and\ \citenamefont
  {Rodriguez}}]{Collins2013nov}%
  \BibitemOpen
  \bibfield  {author} {\bibinfo {author} {\bibfnamefont {L.}~\bibnamefont
  {Collins}}, \bibinfo {author} {\bibfnamefont {J.~I.}\ \bibnamefont
  {Kilpatrick}}, \bibinfo {author} {\bibfnamefont {S.~a.~L.}\ \bibnamefont
  {Weber}}, \bibinfo {author} {\bibfnamefont {A.}~\bibnamefont {Tselev}},
  \bibinfo {author} {\bibfnamefont {I.~V.}\ \bibnamefont {Vlassiouk}}, \bibinfo
  {author} {\bibfnamefont {I.~N.}\ \bibnamefont {Ivanov}}, \bibinfo {author}
  {\bibfnamefont {S.}~\bibnamefont {Jesse}}, \bibinfo {author} {\bibfnamefont
  {S.~V.}\ \bibnamefont {Kalinin}}, \ and\ \bibinfo {author} {\bibfnamefont
  {B.~J.}\ \bibnamefont {Rodriguez}},\ }\href {\doibase
  10.1088/0957-4484/24/47/475702} {\bibfield  {journal} {\bibinfo  {journal}
  {Nanotechnology}\ }\textbf {\bibinfo {volume} {24}},\ \bibinfo {pages}
  {475702} (\bibinfo {year} {2013})}\BibitemShut {NoStop}%
\bibitem [{\citenamefont {Schumacher}\ \emph {et~al.}(2016)\citenamefont
  {Schumacher}, \citenamefont {Miyahara}, \citenamefont {Spielhofer},\ and\
  \citenamefont {Grutter}}]{Schumacher2016apr}%
  \BibitemOpen
  \bibfield  {author} {\bibinfo {author} {\bibfnamefont {Z.}~\bibnamefont
  {Schumacher}}, \bibinfo {author} {\bibfnamefont {Y.}~\bibnamefont
  {Miyahara}}, \bibinfo {author} {\bibfnamefont {A.}~\bibnamefont
  {Spielhofer}}, \ and\ \bibinfo {author} {\bibfnamefont {P.}~\bibnamefont
  {Grutter}},\ }\href {\doibase 10.1103/PhysRevApplied.5.044018} {\bibfield
  {journal} {\bibinfo  {journal} {Phys. Rev. Applied}\ }\textbf {\bibinfo
  {volume} {5}},\ \bibinfo {pages} {044018} (\bibinfo {year}
  {2016})}\BibitemShut {NoStop}%
\bibitem [{\citenamefont {McCulloch}, \citenamefont {Salleo},\ and\
  \citenamefont {Chabinyc}(2016)}]{McCulloch2016jun}%
  \BibitemOpen
  \bibfield  {author} {\bibinfo {author} {\bibfnamefont {I.}~\bibnamefont
  {McCulloch}}, \bibinfo {author} {\bibfnamefont {A.}~\bibnamefont {Salleo}}, \
  and\ \bibinfo {author} {\bibfnamefont {M.}~\bibnamefont {Chabinyc}},\ }\href
  {\doibase 10.1126/science.aaf9062} {\bibfield  {journal} {\bibinfo  {journal}
  {Science}\ }\textbf {\bibinfo {volume} {352}},\ \bibinfo {pages} {1521 }
  (\bibinfo {year} {2016})}\BibitemShut {NoStop}%
\bibitem [{\citenamefont {Bittle}\ \emph {et~al.}(2016)\citenamefont {Bittle},
  \citenamefont {Basham}, \citenamefont {Jackson}, \citenamefont {Jurchescu},\
  and\ \citenamefont {Gundlach}}]{Bittle2016mar}%
  \BibitemOpen
  \bibfield  {author} {\bibinfo {author} {\bibfnamefont {E.~G.}\ \bibnamefont
  {Bittle}}, \bibinfo {author} {\bibfnamefont {J.~I.}\ \bibnamefont {Basham}},
  \bibinfo {author} {\bibfnamefont {T.~N.}\ \bibnamefont {Jackson}}, \bibinfo
  {author} {\bibfnamefont {O.~D.}\ \bibnamefont {Jurchescu}}, \ and\ \bibinfo
  {author} {\bibfnamefont {D.~J.}\ \bibnamefont {Gundlach}},\ }\href {\doibase
  10.1038/ncomms10908} {\bibfield  {journal} {\bibinfo  {journal} {Nat.
  Comms.}\ }\textbf {\bibinfo {volume} {7}},\ \bibinfo {pages} {10908}
  (\bibinfo {year} {2016})}\BibitemShut {NoStop}%
\bibitem [{\citenamefont {Luria}\ \emph {et~al.}(2012)\citenamefont {Luria},
  \citenamefont {Hoepker}, \citenamefont {Bruce}, \citenamefont {Jacobs},
  \citenamefont {Groves},\ and\ \citenamefont {Marohn}}]{Luria2012nov}%
  \BibitemOpen
  \bibfield  {author} {\bibinfo {author} {\bibfnamefont {J.~L.}\ \bibnamefont
  {Luria}}, \bibinfo {author} {\bibfnamefont {N.}~\bibnamefont {Hoepker}},
  \bibinfo {author} {\bibfnamefont {R.}~\bibnamefont {Bruce}}, \bibinfo
  {author} {\bibfnamefont {A.~R.}\ \bibnamefont {Jacobs}}, \bibinfo {author}
  {\bibfnamefont {C.}~\bibnamefont {Groves}}, \ and\ \bibinfo {author}
  {\bibfnamefont {J.~A.}\ \bibnamefont {Marohn}},\ }\href {\doibase
  10.1021/nn300941f} {\bibfield  {journal} {\bibinfo  {journal} {ACS Nano}\
  }\textbf {\bibinfo {volume} {6}},\ \bibinfo {pages} {9392 } (\bibinfo {year}
  {2012})}\BibitemShut {NoStop}%
\bibitem [{\citenamefont {Tennyson}\ \emph {et~al.}(2015)\citenamefont
  {Tennyson}, \citenamefont {Garrett}, \citenamefont {Frantz}, \citenamefont
  {Myers}, \citenamefont {Bekele}, \citenamefont {Sanghera}, \citenamefont
  {Munday},\ and\ \citenamefont {Leite}}]{Tennyson2015sep}%
  \BibitemOpen
  \bibfield  {author} {\bibinfo {author} {\bibfnamefont {E.~M.}\ \bibnamefont
  {Tennyson}}, \bibinfo {author} {\bibfnamefont {J.~L.}\ \bibnamefont
  {Garrett}}, \bibinfo {author} {\bibfnamefont {J.~A.}\ \bibnamefont {Frantz}},
  \bibinfo {author} {\bibfnamefont {J.~D.}\ \bibnamefont {Myers}}, \bibinfo
  {author} {\bibfnamefont {R.~Y.}\ \bibnamefont {Bekele}}, \bibinfo {author}
  {\bibfnamefont {J.~S.}\ \bibnamefont {Sanghera}}, \bibinfo {author}
  {\bibfnamefont {J.~N.}\ \bibnamefont {Munday}}, \ and\ \bibinfo {author}
  {\bibfnamefont {M.~S.}\ \bibnamefont {Leite}},\ }\href {\doibase
  10.1002/aenm.201501142} {\bibfield  {journal} {\bibinfo  {journal} {Adv.
  Energy Mater.}\ ,\ \bibinfo {pages} {1501142}} (\bibinfo {year}
  {2015})}\BibitemShut {NoStop}%
\bibitem [{\citenamefont {Cadena}\ \emph {et~al.}(2016)\citenamefont {Cadena},
  \citenamefont {Sung}, \citenamefont {Boudouris}, \citenamefont
  {Reifenberger},\ and\ \citenamefont {Raman}}]{Cadena2016apr}%
  \BibitemOpen
  \bibfield  {author} {\bibinfo {author} {\bibfnamefont {M.~J.}\ \bibnamefont
  {Cadena}}, \bibinfo {author} {\bibfnamefont {S.~H.}\ \bibnamefont {Sung}},
  \bibinfo {author} {\bibfnamefont {B.~W.}\ \bibnamefont {Boudouris}}, \bibinfo
  {author} {\bibfnamefont {R.}~\bibnamefont {Reifenberger}}, \ and\ \bibinfo
  {author} {\bibfnamefont {A.}~\bibnamefont {Raman}},\ }\href {\doibase
  10.1021/acsnano.5b06893} {\bibfield  {journal} {\bibinfo  {journal} {ACS
  Nano}\ }\textbf {\bibinfo {volume} {10}},\ \bibinfo {pages} {4062 } (\bibinfo
  {year} {2016})}\BibitemShut {NoStop}%
\bibitem [{\citenamefont {Dwyer}\ \emph {et~al.}(2017)\citenamefont {Dwyer},
  \citenamefont {Smieska}, \citenamefont {Tirmzi},\ and\ \citenamefont
  {Marohn}}]{Dwyer2017jul_data}%
  \BibitemOpen
  \bibfield  {author} {\bibinfo {author} {\bibfnamefont {R.~P.}\ \bibnamefont
  {Dwyer}}, \bibinfo {author} {\bibfnamefont {L.~M.}\ \bibnamefont {Smieska}},
  \bibinfo {author} {\bibfnamefont {A.~M.}\ \bibnamefont {Tirmzi}}, \ and\
  \bibinfo {author} {\bibfnamefont {J.~A.}\ \bibnamefont {Marohn}},\ }\href
  {http://github.com/ryanpdwyer/1704-pmkpfm} {\enquote {\bibinfo {title} {Data
  and analysis for ``{V}ector electric field measurement via position-modulated
  {K}elvin probe force microscopy''},}\ }\bibinfo {howpublished} {Available
  from http://github.com/ryanpdwyer/1704-pmkpfm} (\bibinfo {year}
  {2017})\BibitemShut {NoStop}%
\end{thebibliography}%


\begin{thebibliography}{4}%
\makeatletter
\providecommand \@ifxundefined [1]{%
 \@ifx{#1\undefined}
}%
\providecommand \@ifnum [1]{%
 \ifnum #1\expandafter \@firstoftwo
 \else \expandafter \@secondoftwo
 \fi
}%
\providecommand \@ifx [1]{%
 \ifx #1\expandafter \@firstoftwo
 \else \expandafter \@secondoftwo
 \fi
}%
\providecommand \natexlab [1]{#1}%
\providecommand \enquote  [1]{``#1''}%
\providecommand \bibnamefont  [1]{#1}%
\providecommand \bibfnamefont [1]{#1}%
\providecommand \citenamefont [1]{#1}%
\providecommand \href@noop [0]{\@secondoftwo}%
\providecommand \href [0]{\begingroup \@sanitize@url \@href}%
\providecommand \@href[1]{\@@startlink{#1}\@@href}%
\providecommand \@@href[1]{\endgroup#1\@@endlink}%
\providecommand \@sanitize@url [0]{\catcode `\\12\catcode `\$12\catcode
  `\&12\catcode `\#12\catcode `\^12\catcode `\_12\catcode `\%12\relax}%
\providecommand \@@startlink[1]{}%
\providecommand \@@endlink[0]{}%
\providecommand \url  [0]{\begingroup\@sanitize@url \@url }%
\providecommand \@url [1]{\endgroup\@href {#1}{\urlprefix }}%
\providecommand \urlprefix  [0]{URL }%
\providecommand \Eprint [0]{\href }%
\providecommand \doibase [0]{http://dx.doi.org/}%
\providecommand \selectlanguage [0]{\@gobble}%
\providecommand \bibinfo  [0]{\@secondoftwo}%
\providecommand \bibfield  [0]{\@secondoftwo}%
\providecommand \translation [1]{[#1]}%
\providecommand \BibitemOpen [0]{}%
\providecommand \bibitemStop [0]{}%
\providecommand \bibitemNoStop [0]{.\EOS\space}%
\providecommand \EOS [0]{\spacefactor3000\relax}%
\providecommand \BibitemShut  [1]{\csname bibitem#1\endcsname}%
\let\auto@bib@innerbib\@empty
\bibitem [{\citenamefont {Luria}(2011)}]{Luria2011thesis}%
  \BibitemOpen
  \bibfield  {author} {\bibinfo {author} {\bibfnamefont {J.~L.}\ \bibnamefont
  {Luria}},\ }\emph {\bibinfo {title} {Spectroscopic Characterization of Charge
  Generation and Trapping in Third-Generation Solar Cell Materials Using
  Wavelength- and Time-Resolved Electric Force Microscopy}},\ \href@noop {}
  {Ph.D. thesis},\ \bibinfo  {school} {Cornell University} (\bibinfo {year}
  {2011})\BibitemShut {NoStop}%
\bibitem [{\citenamefont {Smieska}(2015)}]{Smieska2015thesis}%
  \BibitemOpen
  \bibfield  {author} {\bibinfo {author} {\bibfnamefont {L.~M.}\ \bibnamefont
  {Smieska}},\ }\emph {\bibinfo {title} {Microscopic Studies of the Fate of
  Charges in Organic Semiconductors: {S}canning {K}elvin Probe Measurements of
  Charge Trapping, Transport, and Electric Fields in p- and n-type Devices}},\
  \href@noop {} {Ph.D. thesis},\ \bibinfo  {school} {Cornell University}
  (\bibinfo {year} {2015})\BibitemShut {NoStop}%
\bibitem [{\citenamefont {Bechhoefer}(2005)}]{Bechhoefer2005Aug}%
  \BibitemOpen
  \bibfield  {author} {\bibinfo {author} {\bibfnamefont {J.}~\bibnamefont
  {Bechhoefer}},\ }\href {\doibase 10.1103/revmodphys.77.783} {\bibfield
  {journal} {\bibinfo  {journal} {Rev. Mod. Phys.}\ }\textbf {\bibinfo {volume}
  {77}},\ \bibinfo {pages} {783} (\bibinfo {year} {2005})}\BibitemShut
  {NoStop}%
\bibitem [{\citenamefont {Hoepker}\ \emph {et~al.}(2011)\citenamefont
  {Hoepker}, \citenamefont {Lekkala}, \citenamefont {Loring},\ and\
  \citenamefont {Marohn}}]{Hoepker2011oct}%
  \BibitemOpen
  \bibfield  {author} {\bibinfo {author} {\bibfnamefont {N.}~\bibnamefont
  {Hoepker}}, \bibinfo {author} {\bibfnamefont {S.}~\bibnamefont {Lekkala}},
  \bibinfo {author} {\bibfnamefont {R.~F.}\ \bibnamefont {Loring}}, \ and\
  \bibinfo {author} {\bibfnamefont {J.~A.}\ \bibnamefont {Marohn}},\ }\href
  {\doibase 10.1021/jp207387d} {\bibfield  {journal} {\bibinfo  {journal} {J.
  Phys. Chem. B}\ }\textbf {\bibinfo {volume} {115}},\ \bibinfo {pages} {14493
  } (\bibinfo {year} {2011})}\BibitemShut {NoStop}%
\end{thebibliography}%
\label{TheEnd}
\end{document}


\title{Supplementary Materials: Vector Electric Field Measurement via Position-Modulated Kelvin Probe Force Microscopy}

\author{Ryan P. Dwyer}
\affiliation{Department of Chemistry and Chemical Biology, Cornell University, Ithaca, New York 14853}

\author{Louisa M. Smieska}
\affiliation{Department of Chemistry and Chemical Biology, Cornell University, Ithaca, New York 14853}

\author{Ali Moeed Tirmzi}
\affiliation{Department of Chemistry and Chemical Biology, Cornell University, Ithaca, New York 14853}

\author{John A. Marohn}
\affiliation{Department of Chemistry and Chemical Biology, Cornell University, Ithaca, New York 14853}

\date{\today}

\maketitle
\thispagestyle{fancy}

\section{Scanned probe microscopy}
\label{sec:spm}

All experiments were performed under vacuum (\SI{1e-6}{\milli\bar}) in a custom-built scanning Kelvin probe microscope \cite{Luria2011thesis,Smieska2015thesis}.
The cantilever (MikroMasch HQ:NSC18/Pt conductive probe) had resonance frequency $f\st{c} = \SI{66401}{\Hz}$, manufacturer-reported spring constant $k\st{c} = \SI{3.5}{\N\per\m}$ and quality factor $Q = 28900$.
Cantilever motion was detected using a fiber interferometer operating at \SI{1490}{\nm} (Corning SMF-28 fiber).
The interferometer light was detected with a \SI{200}{kHz} bandwidth photodetector (New Focus model 2011).

Our FM-KPFM and position-modulated KPFM experiments used the experimental setup of Figure~1a.
A detailed analysis of the experimental setup, feedback loop, and gains can be found in Chapter 5 of Ref.~\citenum{Smieska2015thesis}.

\begin{figure}
\cgraphics[width=6in]{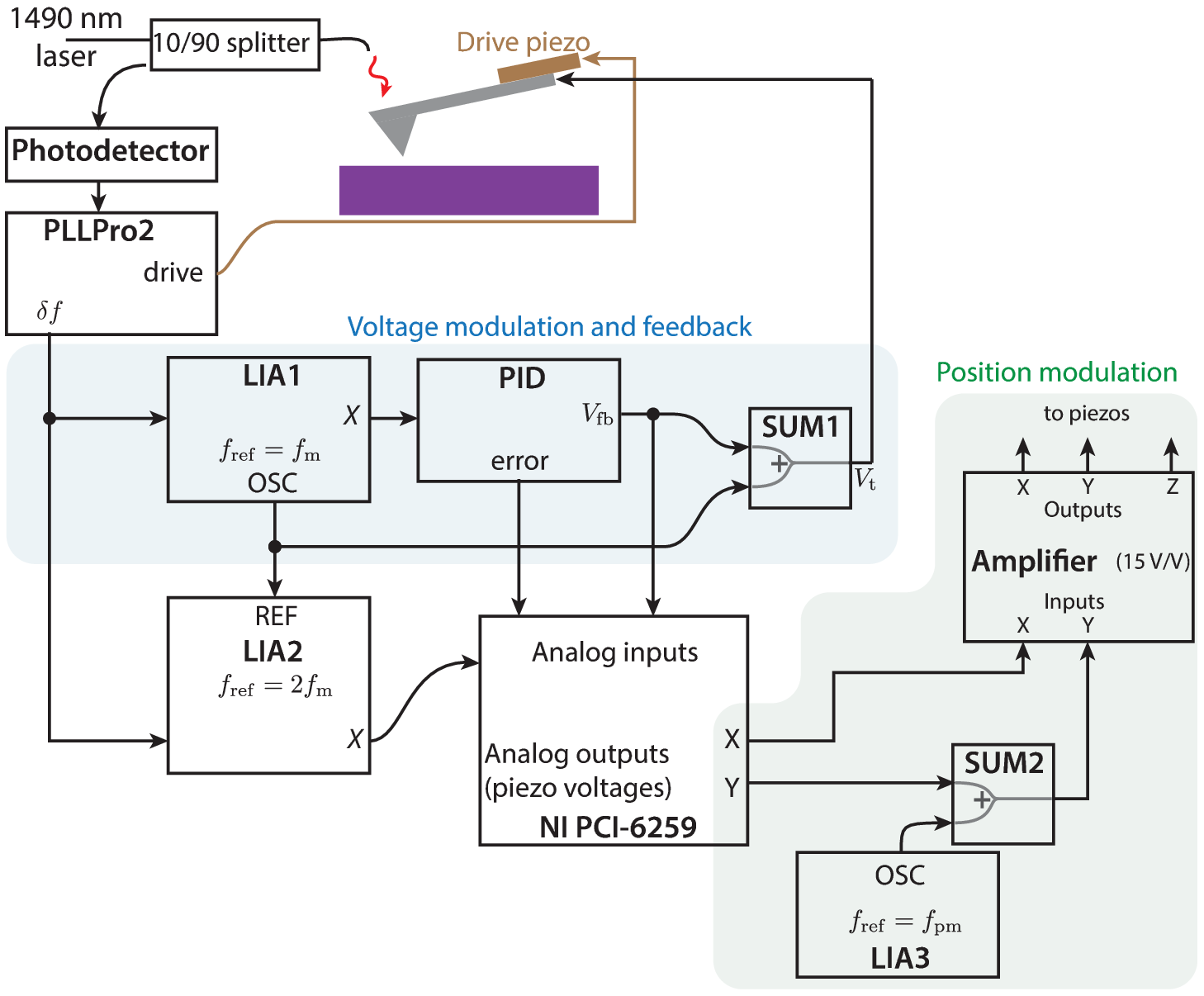}
\caption{A block diagram showing the hardware used to perform the voltage modulation and position modulation.}
\label{fig:block-diagram}
\end{figure}

Figure~\ref{fig:block-diagram} shows the hardware used to perform the voltage modulation and position modulation in more detail.
The cantilever was driven using a commercial phase locked loop (PLL) cantilever controller (RHK Technology, PLLPro2 Universal AFM controller), with PLL loop bandwidth $\SI{1}{\kilo\hertz}$ (PLL feedback loop integral gain $I = \SI{2.5}{\per\Hz}$, proportional gain $P = \SI{-10}{degrees\per\Hz}$).
The PLL frequency shift $\delta f$ was input to a Stanford Research Systems SR830 lock-in amplifier (LIA1, Fig.~\ref{fig:block-diagram}) operating at the voltage-modulation frequency $f\st{m} = \SI{160}{\Hz}$.
We set the LIA1 filter time constant to $\SI{10}{\ms}$, the filter slope to $\SI{6}{\decibel\per\octave}$ and turned the synchronous filter setting on.
To auto-phase LIA1, the time constant was increased to $\mathrm{TC} = \SI{300}{\ms}$ and a \SI{1}{\V} DC offset was applied to the tip ($V\st{ts} - \phi = \SI{1}{\V} + V\st{m}\sin(2\pi f\st{m} t)$).
The $X$-channel LIA1 output, proportional to the cantilever frequency shift component at the modulation frequency $\delta f(f\st{m})$, was used as the measurement input to a proportional-integral-derivative (PID) controller (Stanford Research Systems SIM960).

We set the PID controller integral gain to $I = \SI{50}{\radian\per\s}$ and derivative gain to $D = \SI{5e-5}{\s}$. 
Before each KPFM scan, we set the proportional gain by increasing the gain until the feedback loop became unstable ($P = P\st{unstable}$) and then reducing the gain to $P = P\st{unstable}/3$, corresponding to a gain margin of \num{3}.
Typically, the resulting proportional gain was $P = 0.1$ to $0.6$.
With this gain margin criteria, the overall feedback loop bandwidth was $b \sim \SI{40}{\Hz}$, measured using the network analyzer function on a Digilent Analog Discovery USB Oscilloscope and Logic Analyzer (see Figures~5.12 and 5.13 in Ref.~\citenum{Smieska2015thesis}).
The proportional gain was set over the higher capacitance source and drain electrodes; the overall bandwidth was lower over the transistor channel ($b = 20$ to \SI{30}{\Hz}).

The applied tip voltage was the sum of the $f\st{m}$-LIA oscillator output voltage (OSC on LIA1, Fig.~\ref{fig:block-diagram}) and the PID output voltage $V\st{fb}$.
We set the LIA1 oscillator rms-voltage to $V\st{rms} = V\st{m} / \sqrt{2} = \SI{1.5}{\V}$.
A home-built summing circuit (Analog Devices AD711 operational amplifier; SUM1 in Fig.~\ref{fig:block-diagram}) was used to add the two voltages.

A second lock-in amplifier (PerkinElmer Signal Recovery 7265; LIA2 in Fig.~\ref{fig:block-diagram}) was used to track the component of the cantilever frequency shift at twice the voltage-modulation frequency $\delta f(2\st{m})$.
This frequency shift component is proportional to the tip-sample capacitance derivative $C''(d)$.
We set the LIA2 filter time constant $\mathrm{TC} = \SI{50}{\ms}$ and the filter slope to $\SI{6}{\decibel\per\octave}$.

The sample position was modulated using the oscillator of a third lock-in amplifier (PerkinElmer Signal Recovery 7265, LIA3 in Fig.~\ref{fig:block-diagram}) set to the position-modulation frequency $f\st{pm}$.
To apply the position modulation along the KPFM-scan axis (as in Figure~2), a home-built summing circuit (Analog Devices AD711 operational amplifier; SUM2 in Fig.~\ref{fig:block-diagram}) was used to add the position-modulation sinusoid to the KPFM ramp signal (typically \num{-5} to \SI{5}{\V}).
The summing circuit output was amplified by $\SI{15}{V}/\si{V}$ using a Piezomecanik 350 amplifier (Amplifier in Fig.~\ref{fig:block-diagram}).

The tip-sample distance $h$ was set by tapping the surface (60 percent amplitude set point, cantilever zero-to-peak amplitude $A = \SI{50}{\nm}$ away from the surface) over the source or drain electrode then retracting the $z$-piezo by $\Delta z = h - 0.6A$ so that the mean tip-sample distance was $h$.

For PM-KPFM and KPFM line scans, the PID feedback voltage $V\st{fb} = \phi\st{meas}$, PID error signal, and the $X$-channel of LIA2, proportional to the tip-sample capacitance, were digitized at a sampling rate $f\st{s} = \SI{8.192}{\kHz}$ using custom LabVIEW data acquisition code and a National Instruments PCI-6259.
The full $\phi\st{meas}$, PID error signal, and tip-capacitance transients were saved for further analysis.

\begin{figure}
\includegraphics{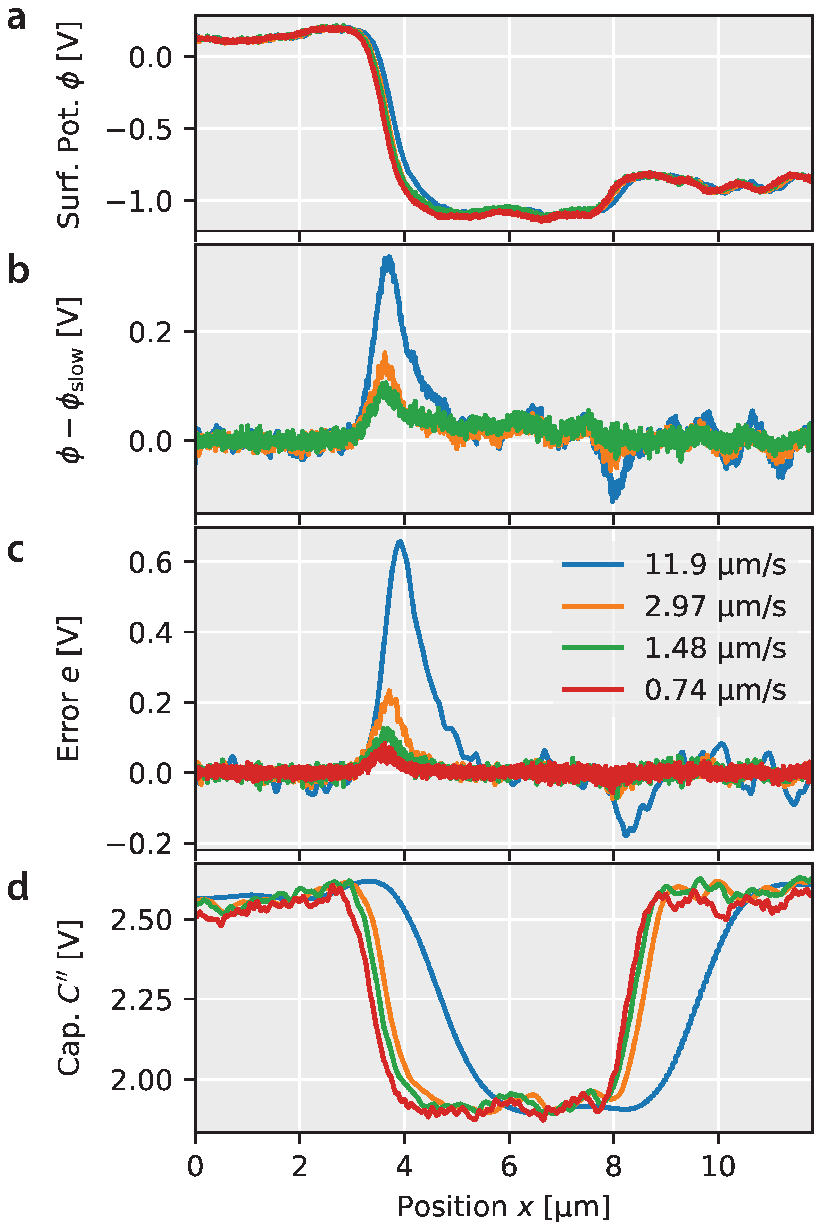}
\caption{
Increasing scan speed causes surface potential errors.
The scan velocity is given in the legend.
(a) Measured surface potential $\phi\st{meas} = V\st{fb}$.
(b) For the faster scan speeds $v$, the difference in surface potential compared to the slowest line scan: $\phi(v) - \phi(\SI{0.74}{\um\per\s})$.
(c) The $X$-channel of the LIA operating at $f\st{m}$, which is the feedback loop error signal, $X_{f\st{m}} \propto \delta f(f\st{m})$.
(d) The $X$-channel of the LIA operating at $2f\st{m}$, which is proportional to the capacitance derivative $C''$.
Experimental parameters: Tip-sample separation $h = \SI{200}{\nm}$, $V\st{m}/\sqrt{2} = \SI{1.5}{\V}$, transistor voltages $V\st{S} = 0$, $V\st{D} = \SI{-1}{\V}$, and $V\st{G} = \SI{-10}{\V}$.
}
\label{fig:scan-speed}
\end{figure}

\section{Scan speed}
\label{sec:scan-speed}

Figure~\ref{fig:scan-speed} shows that with increasing scan speed, the feedback loop could not track the abrupt changes in the sample's surface potential.
We scanned across the transistor channel (Fig.~2) at a series of increasing scan speeds.
We saw clear differences in the measured surface potential profile (Fig.~\ref{fig:scan-speed}a) as the scan speed was increased from $v = \SI{0.74}{\um\per\s}$ to \SI{11.9}{\um\per\s}. 
Fig.~\ref{fig:scan-speed}b highlights these differences by plotting the difference between the higher speed linescans $\phi(x, v=\SI{1.48}{\um\per\s}\ldots)$ and the slowest linescan $\phi(x, v=\SI{0.74}{\um\per\s})$.
The differences are consistent with increasing error $\phi\st{error} = \phi - \phi\st{meas}$ caused by the feedback loop responding too slowly to keep $V\st{fb} = \phi$ as the tip scans across the edge of the transistor channel.

This explanation is supported by the corresponding increase in the PID error $e$ shown in Fig.~\ref{fig:scan-speed}c.
The error $e$ is the $X$-channel of the phased lock-in amplifier operating at the modulation frequency $f\st{m}$, which is related to the surface potential error $\phi\st{error}$:
$X_{f\st{m}} \propto \delta f(f\st{m}) =  -f\st{c} C'' V\st{m} \phi\st{error} /(2k\st{c})$.
The capacitance data show the slow response of the lock-in amplifier operating at $2f\st{m}$ (Fig.~\ref{fig:scan-speed}d).

Based on this data, in the manuscript we chose a tip velocity $v = \SI{0.37}{\um\per\s}$ for pm-FM-KPFM and FM-KPFM linescans.
For FM-KPFM images, we chose a tip velocity $v = \SI{1.5}{\um\per\s}$ (\SI{8}{\s}/\text{line}) so that an entire $128$ by $128$ image could be collected in \SI{17}{\min}.


\clearpage

\section{Data analysis}
\label{sec:data-analysis}

Using the full surface potential transients, the Figure~1b data analysis was performed in Python.
The raw surface potential $\phi\st{meas} = \Phi$ versus time transient was filtered using the finite-impulse-response filter $H\st{pm}$ to produce an estimate of the sample's surface potential $\phi$:
\begin{equation}
\phi = H\st{pm} \ast \Phi,
\end{equation}
where $\ast$ denotes discrete time convolution.
We discarded the portion of $\phi$ where the filter $H\st{pm}$ did not overlap fully with the data $\Phi$.
We associated each data point $\phi_{i}$ with the position along the scan axis $x_i = v t_i$, where $v$ is the tip velocity, $i = 0,1 \ldots N-1$ is the index, $N = \num{262144}$ the number of data points collected, the time $t_{i} = i / f\st{s}$, and the sampling rate $f\st{s} = \SI{8.192}{\kHz}$.

\begin{figure}
\includegraphics{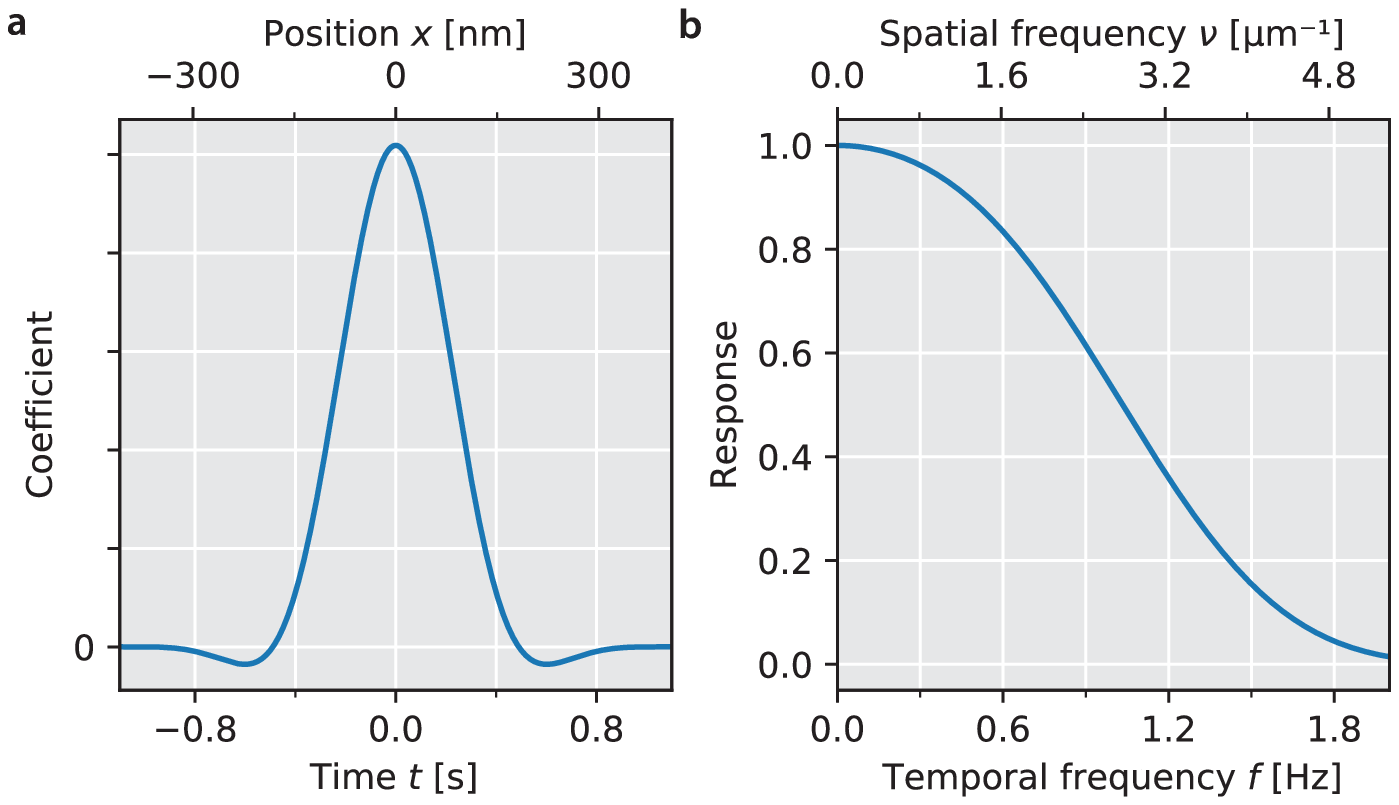}
\caption{
The software lock-in amplifier filter.
(a) The finite impulse response lock-in amplifier filter $H\st{pm}$ in the time domain.
The upper position axis shows $x = v t$, with $v = \SI{0.37}{\um\per\s}$ the cantilever tip velocity along the scan axis.
(b) $H\st{pm}$ in the frequency domain. The upper spatial frequency axis shows $\nu = f/v$, with $v = \SI{0.37}{\um\per\s}$ the cantilever tip velocity along the scan axis.
}
\label{fig:filter}
\end{figure}

To calculate the electric field via numerical differentiation ($E = - \partial \phi / \partial x$), we used a 2nd order central difference approximation to compute the derivative of the potential
\begin{equation}
E_{i} = - \frac{(\phi_{i+1} - \phi_{i-1})}{2\Delta x} = - \frac{(\phi_{i+1} - \phi_{i-1})}{2 v / f\st{s}}.
\label{eq:E-nd}
\end{equation}
where $\Delta x$ is difference in position between adjacent data points: $\Delta x = x_i - x_{i-1} = v / f\st{s}$.

\begin{figure}
\includegraphics{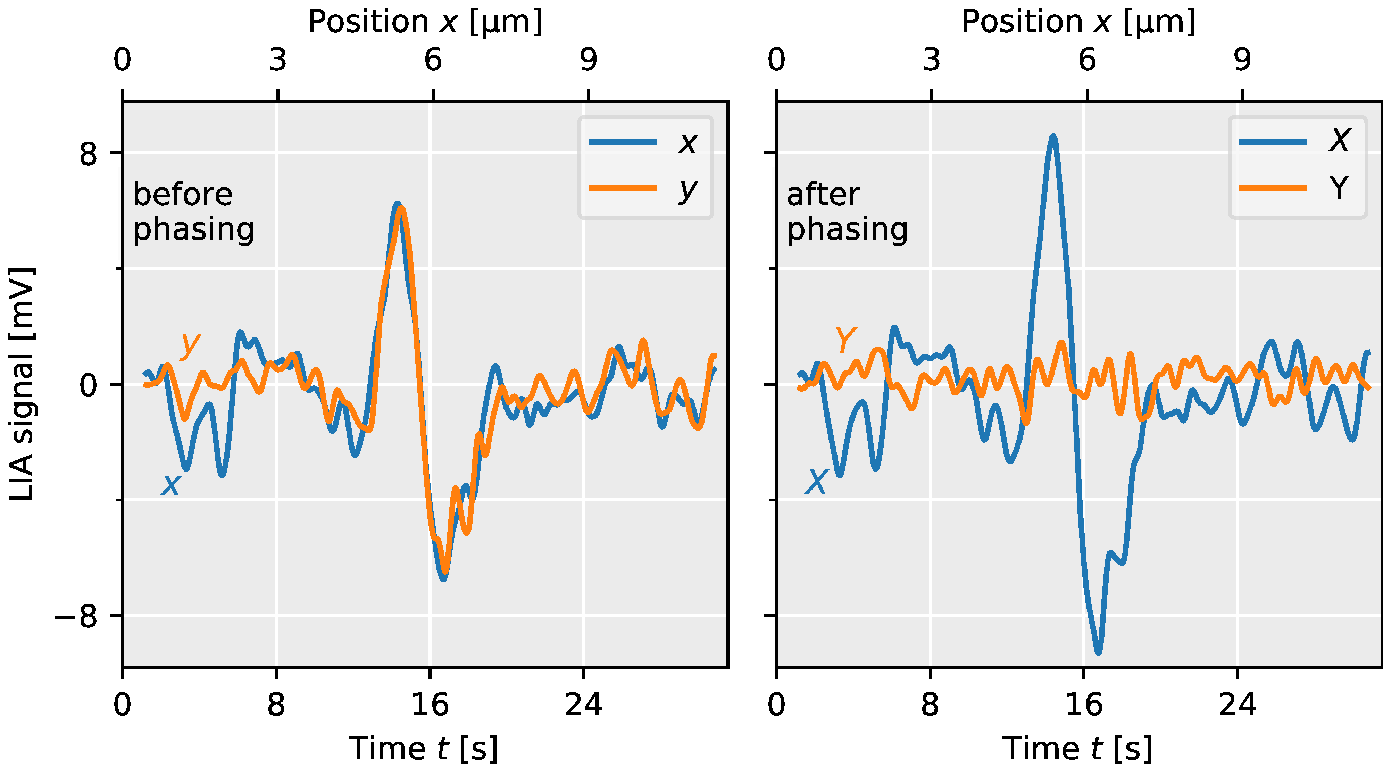}
\caption{
The software lock-in amplifier output channels (a) before phasing and, (b) after phasing.
The best-fit parameters (Eq.~\ref{eq:phase}) were $\Delta f = \SI{4.8}{\milli\Hz}$ and $\theta_0 = \SI{0.26}{\radian}$.
}
\label{fig:phase}
\end{figure}

To calculate the electric field using the modulation component, we processed the raw surface potential $\phi\st{meas} = \Phi$ with a software lock-in amplifier filter (Fig.~\ref{fig:filter}).
First, the complex lock-in amplifier signal $z$ was generated using
\begin{equation}
z =  H\st{pm} \ast \Big (\exp(-2\pi j \, f\st{pm} t) \times \Phi \Big),
\end{equation}
where $H\st{pm}$ is the finite-impulse-response filter, $j = \sqrt{-1}$, $f\st{pm} = \SI{4.5}{\Hz}$ is the position modulation frequency, and $t$ is a vector of time data $t = (0, 1, \ldots N-1)/f\st{s}$.
We discarded the portion of $z$ where the filter $H\st{pm}$ did not overlap fully with the data (about \SI{1.1}{\s} at the beginning and end of the data set).
From $z$, we calculated the signal's amplitude $A = \abs{z}$ and phase $\theta = \arg{z}$.
The real ($x = \Re{z}$) and imaginary ($y = \Im{z}$) components of $z$ are shown in Figure~\ref{fig:phase}a.

The lock-in amplifier was phased by minimizing the signal in the out-of-phase channel.
When phasing the lock-in amplifier, we also adjusted the lock-in amplifier reference frequency.
This procedure corrects for any correct slow drift in the phase.
We optimize using the cantilever amplitude and phase as follows:
\begin{equation}
\min_{\Delta f, \theta_0} \, \sum_{i}^{N} A_{i}^{2} \left [ \,  \Bigg \lvert \lvert \underbrace{\theta_{i} - 2 \pi \Delta f t_{i} + \theta_0}_{\Delta \theta} \rvert - \frac{\pi}{2} \Bigg \rvert - \frac{\pi}{2} \right ]^{2},
\label{eq:phase}
\end{equation}
where the frequency $\Delta f$ is a correction to the reference frequency (typically $0$ to $\SI{10}{\mHz}$) and $\theta_0$ is the lock-in amplifier phase.
The under-braced term is the ordinary phase difference; the rest of the bracketed expression accounts for phase reversals caused by a change in sign of $E\st{mod}$.
With this correction, a phase difference $\Delta \theta = \pi$ contributes $0$ to the sum, since $\Delta \theta = \pi$ corresponds to a signal of the opposite sign.
After performing the minimization, the phased lock-in amplifier output is
\begin{equation}
Z = X + Y j =  z \, \exp \big(-2\pi j \Delta f \, t - \theta_0 \, j \big),
\end{equation}
plotted in Figure~\ref{fig:phase}b.
The electric field along the modulation direction was $E\st{pm} = X / A\st{pm}$, where $A\st{pm}$ was the zero-to-peak amplitude of the position modulation.

\section{Transfer function sensitivity analysiss}
\label{sec:correction-term}
The surface potential was measured using the KPFM feedback loop.
The position modulation causes a modulation in the feedback loop transfer function, which can be seen by writing the closed-loop transfer function $\hat{H}$ (the Fourier transform of the impulse response function in Eq.~3) can be written in terms of the open-loop transfer function $\hat{L}$:
\begin{equation}
\hat{H}(f) = \frac{\hat{L}(f)}{1+\hat{L}(f)}
\end{equation}
The open-loop transfer function $\hat{L}(f)$ is the product of the transfer functions of the individual loop components:
\begin{equation}
\hat{L}(f) =  \hat{H}\st{PID}(f) \hat{H}\st{LIA}(f) \hat{H}\st{PLL}(f) \frac{f\st{c}}{2 k\st{c}} C''(\vec{r}) V\st{m}.
\end{equation}
The position modulation can cause an oscillation in the measured surface potential due an oscillation in the tip-sample capacitance $C''(\vec{r})$ related to the modulation.
We expect the modulation to be small because we use position-modulation amplitudes that are small compared to the sample's expected feature size.
For example, in our sample, the maximum fractional change in $C''$ would occur scanning across the edge of the transistor channel.
The maximum fractional rate of change of $\partial (C''/C''_0) / \partial x$ is $\SI{0.6}{\per\um}$ (Fig.~\ref{fig:scan-speed}).
At our position-modulation amplitude $A\st{pm} = \SI{30}{\nm}$, this corresponds to a worst-case fractional change of 1.8 percent.
By using a closed-loop technique to measure the surface potential, we attenuate any contribution to the measured signal by a factor related to the transfer function sensitivity $d\hat{H} = d\hat{L} / (1+\hat{L})^2$ \cite{Bechhoefer2005Aug}.
Inverting the measured transfer function $\hat{H}(f)$, we estimate that $\hat{L}(f=\SI{4.5}{\Hz}) = 0.6 - 5.9 j$.
The \SI{4.5}{\Hz} lock-in amplifier signal is affected by disturbances to the lock-in amplifier $X$-channel: $\Re{H^{*}(f) \, d\hat{H}(f)}/\abs{H(f)}$.
The contribution of any fractional change in capacitance to the measured signal is attenuated by the closed-loop feedback by a factor of 20.
Therefore the largest possible contribution of the change in $C''$ to the measured electric field signal would be a 0.1 percent change in the magnitude of the electric field measured.


\section{Slow axis KPFM electric field measurement}
\label{sec:slow-axis-E}
The KPFM surface potential shown in Fig.~4a was not filtered.
The electric field along the scan axis (Fig.~4b) was calculated using Equation~\ref{eq:E-nd} with $\Delta x = \SI{90}{\nm}$.
Before calculating the electric field along the slow-scan axis (dark squares), the KPFM image was filtered along the $y$-axis using a filter designed using the same procedure as $H\st{pm}$, with the same spatial bandwidth $\nu_y^{\text{3-dB}} = \SI{2.2}{\per\um}$.
The unfiltered $E_y$ determined from FM-KPFM had a spatial bandwidth in $y$ of $\nu_y^\text{3-dB} = \SI{4.4}{\per\um}$ set by the bandwidth of the numerical derivative filter: $\sin(2\pi \Delta y \nu_y) / (2 \pi \Delta y \nu_y)$.
The pm-FM-KPFM bandwidth along the $y$-axis $\nu_y^\text{3-dB} = \SI{8.6}{\per\um}$ is determined by the size of the position-modulation amplitude $A\st{pm}$: $J_1(2 \pi A\st{pm} \nu_y) / (\pi A\st{pm} \nu_y)$ where $J_1$ is a Bessel function of the first kind.

\section{Modulating and scanning in perpendicular directions}

\begin{figure}
\includegraphics{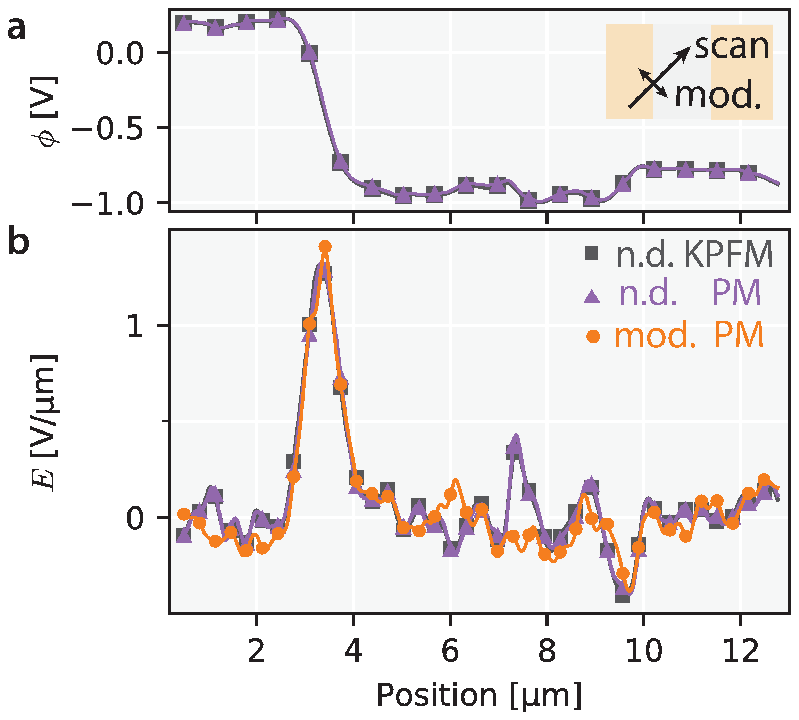}
\caption{Comparison of electric fields measured using pm-FM-KPFM and KPFM. (a) Surface potential measured using KPFM and position-modulated KPFM. The scan and modulation are at $+\num{45}$ and $\num{-45}$ degrees relative to the $x$-axis respectively (inset).
(b)
Comparison of the electric fields measured using FM-KPFM and position-modulated KPFM.
Numerical derivative of the FM-KPFM surface potential (squares), numerical derivative of the pm-FM-KPFM surface potential (triangles), and the modulation component of the pm-FM-KPFM surface potential (circles).
Experimental parameters: position-modulation amplitude $A\st{pm} = \SI{30}{\nm}$, frequency $f\st{pm} = \SI{4.5}{\Hz}$, tip velocity $v=\SI{414}{\nm\per\s}$.
}
\label{fig:pm-kpfm-perp}
\end{figure}

Figure~\ref{fig:pm-kpfm-perp} shows the new position-modulated KPFM measurement applied to measure perpendicular components of the electric field simultaneously.
We scanned across the transistor channel (Fig.~2) at an angle of $+45$ degrees relative to the $x$-axis and applied a position modulation perpendicular to the scan axis (Fig.~\ref{fig:pm-kpfm-perp}a inset).
In order to apply the modulation at an angle, an additional summing circuit SUM3 (AD711 op-amp) was used to add the modulation to the $x$-axis as well as the $y$-axis:
\begin{align}
V\st{x-piezo}& = G\st{amp} \underbrace{(V\st{x,ramp}(t) + V\st{OSC}(t))}_{\text{SUM3}} \\
V\st{y-piezo}& = G\st{amp} \underbrace{(V\st{y,ramp}(t) + V\st{OSC}(t)) }_{\text{SUM2}}
\end{align}
where the sums are labeled as in Fig.~\ref{fig:block-diagram} and $V\st{x,ramp}$ and $V\st{y,ramp}$ are the $X$ and $Y$ voltage outputs from the NI PCI-6259.

While the scan and modulation signals measure perpendicular components of the electric field in the Fig.~\ref{fig:pm-kpfm-perp}a experiment, both observe the same electric field drop at the transistor contacts because there the electric field was entirely along the $x$-axis (Fig.~2c).
At the contacts, the electric field along the scan direction is $E\st{scan} = E_{x} \cos (\SI{45}{\degree}) = E_{x} /\sqrt{2}$.
Likewise, the electric field along the modulation direction is $E\st{mod} = E_{x} \cos (-\SI{45}{\degree}) = E_{x}/\sqrt{2}$.
Fig.~\ref{fig:pm-kpfm-perp}b shows that the position-modulated measurement tracks the electric field as measured by other KPFM measurements.
At the source contact, the measured electric field agrees with the electric field measured in Fig.~2d.

\section{Noise in FM-KPFM}

\begin{figure}
\includegraphics[width=\textwidth]{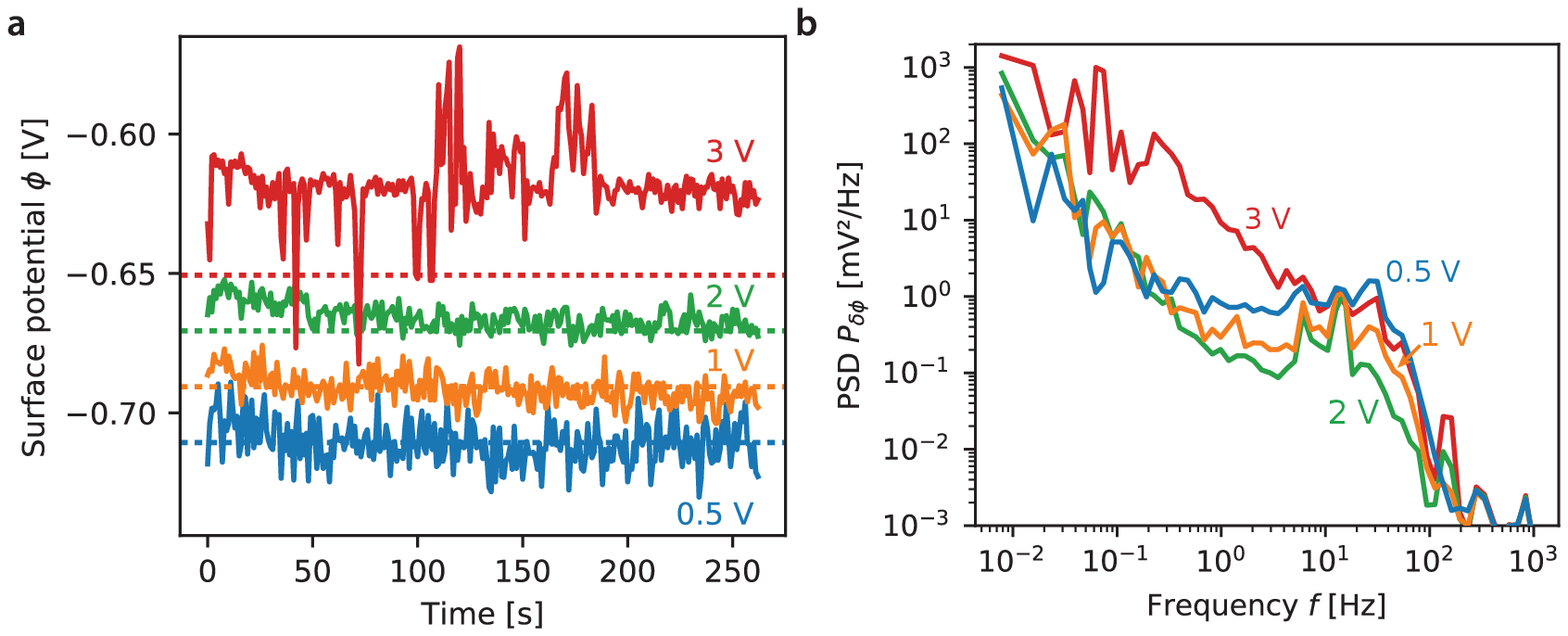}
\caption{
FM-KPFM surface potential noise.
(a) Surface potential \emph{versus} time measured through the FM-KPFM feedback loop for different modulation voltages $V\st{m}/\sqrt{2} = V\st{rms} = \num{0.5}$, \num{1.0}, \num{2.0} and \SI{3}{\V}.
(b)
Power spectral density of surface potential fluctuations $P_{\delta \phi}$ calculated from the data in (a).
}
\label{fig:pm-kpfm-noise}
\end{figure}

We measured the surface-potential noise under the optimized conditions used in this experiment.
We collected surface-potential-\latin{versus}-time data over the drain electrode ($V\st{D} = \SI{-1}{\V}$) with the sample stationary (Fig.~\ref{fig:pm-kpfm-noise}a).
The surface potential had some high-frequency noise plus slow drift on the seconds and longer time scale.

To better illustrate the surface potential noise \emph{versus} frequency, Fig.~\ref{fig:pm-kpfm-noise}b plots the power spectral density of surface potential fluctuations calculated from the time domain data in Fig.~\ref{fig:pm-kpfm-noise}a.
The slow drift caused increased spectral density below \SI{1}{\Hz}.
For rms-modulation voltages $\SI{2}{\V}$ or less, the low-frequency noise was not cantilever frequency noise, because the contribution of cantilever frequency noise $P_{\delta f}$ to surface potential noise is inversely proportional to the modulation voltage $V\st{m}$.
There was a small region from approximately \num{1} to \SI{10}{\Hz} where the surface potential noise was minimized.
For rms-modulation voltages $\SI{2}{\V}$ or less, the surface potential noise in the majority of this region was inversely proportional to the modulation voltage $V\st{m}$.
This observation is consistent with voltage noise arising from position-detection noise \cite{Hoepker2011oct}.
The power spectral density of surface potential fluctuations exhibited a roll-off between \num{10} and \SI{20}{\Hz} due to the feedback-network filter, as expected.

The electric field calculated by numerically differentiating the FM-KPFM image along the slow-scan axis incorporates surface potential noise at temporal frequencies near the inverse line-scan time $(\SI{8}{\s}/\text{line})^{-1} \sim \SI{0.1}{\Hz}$, where noise is increased by the slow drift.
The pm-FM-KPFM measurement incorporates surface potential noise at frequencies near $f\st{pm} = \SI{4.5}{\Hz}$, where overall surface potential noise is near a minimum.

\clearpage

\begin{figure}
\includegraphics{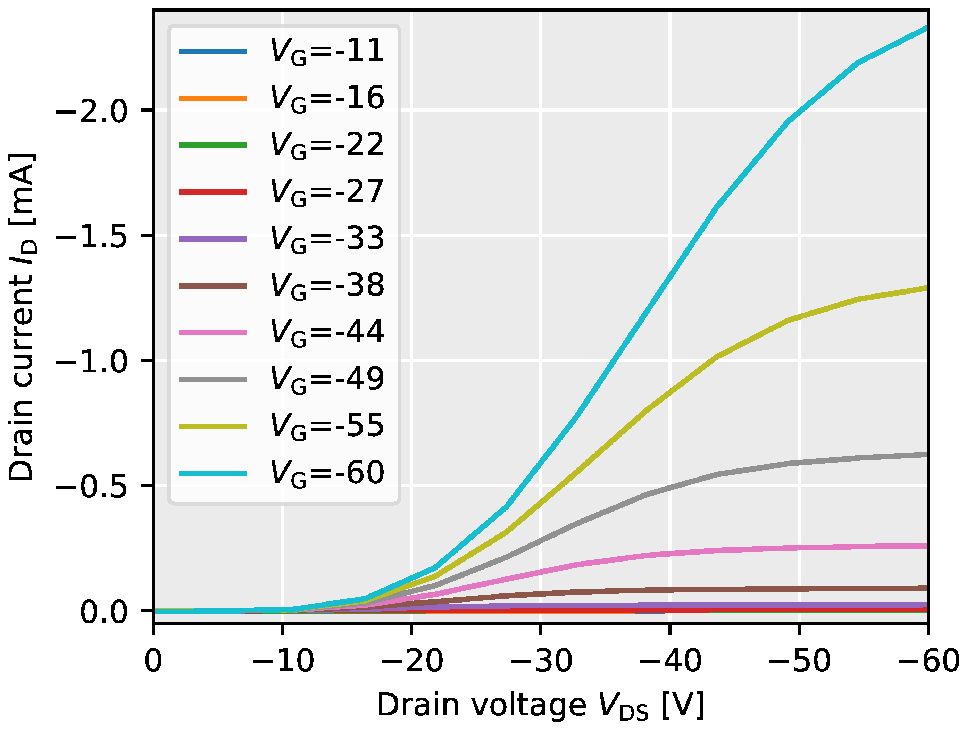}
\caption{Current-voltage curves collected on the DPh-BTBT transistor studied in this manuscript.}
\label{fig:iv-curves}
\end{figure}

\bibliography{Dwyer}